%% file: paper.tex
\documentclass[article,pageno]{jpaper}[]

\usepackage{algorithmic, algorithm, amsmath}
\usepackage[normalem]{ulem}
\usepackage{xcolor}
\usepackage{flushend}
\usepackage{url}

\usepackage{breakurl}
\usepackage{authblk}

\usepackage[normalem]{ulem}
\def\Section {\S}

\author{Abel Souza}
\author{Noman Bashir}
\author{Jorge Murillo}
\author{Walid Hanafy}
\author{Qianlin Liang}
\author{David Irwin}
\author{\\Prashant Shenoy}
\affil{University of Massachusetts Amherst}

\begin{document}

\title{Ecovisor: A Virtual Energy System for Carbon-Efficient Applications\thanks{Paper to appear at ASPLOS 2023.}}

\date{}
\maketitle

\begin{abstract}
\input{abstract}

\end{abstract}

\thispagestyle{empty}

\section{Introduction}
\label{sec:introduction}
\input{introduction}

\section{Motivation and Background}
\label{sec:background}
\input{background}

\section{Ecovisor Design}
\label{sec:design}
\input{design}

\section{Prototype Implementation}
\label{sec:implementation}
\input{implementation}

\section{Optimizing Carbon-Efficiency}
\label{sec:evaluation}
\input{evaluation.new}

\section{Related Work}
\input{related}

\section{Conclusion}
\input{conclusion}

\section*{Acknowledgements}
We thank ASPLOS reviewers for their valuable comments and electricityMap for their carbon intensity dataset. This research is supported by NSF grants 2213636, 2105494, 2021693, 2020888, as well as VMware.

\bibliographystyle{plain}
\bibliography{paper}

\end{document}

%% file: abstract.tex
Cloud platforms' rapid growth is raising significant concerns about their carbon emissions. To reduce emissions, future cloud platforms will need to increase their reliance on renewable energy sources, such as solar and wind, which have zero emissions but are highly unreliable. Unfortunately, today's energy systems effectively mask this unreliability in hardware, which prevents  applications from optimizing their carbon-efficiency, or work done per kilogram of carbon emitted.  To address this problem, we design an ``ecovisor,'' which virtualizes the energy system and exposes software-defined control of it to applications.  An ecovisor enables each application to handle clean energy's unreliability in software based on its own specific requirements.  We implement a small-scale ecovisor prototype that virtualizes a physical energy system to enable software-based application-level i) visibility into variable grid carbon-intensity and renewable generation and ii) control of server power usage and battery charging/discharging.  We evaluate the ecovisor approach by showing how multiple applications can concurrently exercise their virtual energy system in different ways to better optimize carbon-efficiency based on their specific requirements compared to a general system-wide policy.

%% file: introduction.tex
Cloud platforms are growing exponentially, and have been for some time, with a recent analysis estimating a 6$\times$ increase in their capacity from 2010-2018, or roughly a 22.4\% increase per year~\cite{masanet}.  This hyperscale growth is being driven by the continual development of new and useful, but often computationally-intensive, applications, particularly in artificial intelligence (AI)~\cite{ml-compute-demand}.   As they have grown, cloud platforms have aggressively optimized their energy-efficiency, e.g., by reducing their power usage effectiveness (PUE)\footnote{The PUE is the ratio of total data center power to server power.}  to near the optimal value of $1$~\cite{google-pue,uptime}, to mitigate large increases in their energy consumption and cost. 

However, further improving energy-efficiency is becoming increasingly challenging, as it is already highly optimized. Thus, continued growth in cloud capacity will likely result in much larger increases in energy consumption moving forward.  Of course, this energy growth is also increasing cloud platforms' carbon and greenhouse gas (GHG) emissions, which are causing climate change~\cite{guardian,ipcc-report}.  The negative environmental effects of the cloud's hyperscaler growth have begun to receive significant attention.  As a result, all the major cloud providers have announced aggressive goals for reducing, and ultimately eliminating, their platforms' carbon emissions over the next decade, while acknowledging that many of the technologies necessary to achieve these sustainability goals have yet to be developed~\cite{amazon-carbon-neutral,vmware-carbon,google-carbon-free,facebook-carbon-neutral,microsoft-carbon-negative}.

Reducing cloud platforms' carbon emissions will require them to power their cloud and edge data centers using cleaner ``lower-carbon'' energy sources.  
A distinguishing characteristic of clean energy is its unreliability:  it is intermittent and not available in unlimited quantities at any single location all the time.   
Notably, clean energy's unreliability manifests itself in two distinct ways within our current energy system:  i) the unreliability of renewable power generation and ii) the volatility of grid power's carbon-intensity. 
In the former case, the power generated by zero-carbon renewable energy sources, primarily solar and wind, at any location is unreliable because it varies based on changing environmental conditions.  
In the latter case, the carbon-intensity of grid power --- in kg$\cdot$CO$_2$ equivalent per watt (W) --- is volatile because it varies based on the carbon emissions of the different types of generators the grid uses to satisfy its variable demand. 
As we discuss, both forms of unreliability are important to consider in reducing cloud platforms' carbon emissions. 

Compared to other industries, computing is uniquely well-positioned to reduce its carbon emissions by transitioning to cleaner energy sources, despite their unreliability, for numerous reasons.   Most importantly, computation often has significant spatial, temporal, and performance flexibility, which enables execution location, time, and intensity shifting to better align with the availability of low-carbon grid power and zero-carbon renewable power~\cite{greenslot,parasol,sharma:asplos11}.  In addition, computation can also leverage numerous software-based fault-tolerance techniques, including checkpointing, replication, and recomputation, to continue execution despite unexpected variations in the availability of low-carbon energy, which may require throttling or shutting down servers~\cite{yank}.  

Unfortunately, today's cloud applications cannot leverage the unique combination of advantages above to optimize their \emph{carbon-efficiency}, or work done per kilogram (kg) of carbon (and other GHGs) emitted, because current energy systems effectively mask clean energy's unreliability from them in hardware.   That is, energy systems have traditionally exposed a \emph{reliability abstraction} -- the abstraction of a reliable supply of power on demand up to some maximum -- to electrical devices, including servers, via their electrical socket interface.  In many cases, the energy system now includes a connection to not only the grid, but also an increasingly rich local energy system that may include substantial energy storage, e.g., batteries~\cite{dcd-article}, and co-located renewable energy sources, e.g., wind and solar~\cite{maiden}.   Since energy systems hide their increasing complexity behind the reliability abstraction, they provide applications no control of, or visibility into, the characteristics of their energy supply, i.e., its consumption, generation, or carbon emissions. Thus, applications cannot optimize carbon-efficiency by regulating their power usage to respond to changes in grid power's carbon-intensity and renewable power's availability.

To address the problem, this paper presents the design and implementation of an \emph{ecovisor}---a 
software system that exposes software-defined control of a virtual energy system directly to applications.  An ecovisor is akin to a hypervisor but virtualizes the energy system of computing infrastructure instead of virtualizing the computing resources of a single server. Importantly, an ecovisor enables applications to handle clean energy's unreliability within their software stack based on their own specific characteristics, performance requirements, and sustainability goals by leveraging one or more dimensions of software flexibility and software-based fault-tolerance. Ecovisors also enable applications to exercise software-based control of their virtual energy system to mitigate clean energy's unreliability.  Specifically, instead of temporally or spatially shifting their computing workload, applications can control their virtual battery to temporally shift their clean energy usage---by storing renewable or low-carbon grid energy when it is available for later use. 

In some sense, our approach extends the end-to-end principle~\cite{endtoend} to the energy system by i) recognizing that the energy system's current reliability abstraction prevents designing carbon-efficient applications, and ii) addressing the problem by pushing control of the energy system from hardware into software.   Our approach is also inspired by the exokernel argument from operating systems that advocates delegating resource management to applications~\cite{exokernel,mesos}.  Our ecovisor extends this approach by delegating not only resource management to applications, but also the energy (and carbon) that powers those resources. Our hypothesis is that exposing software-defined visibility and control of a virtualized energy system enables applications to better optimize carbon-efficiency based on their specific characteristics and requirements compared to a general system policy.  In evaluating our hypothesis, this paper makes the following contributions.

\noindent {\bf Virtualizing the Energy System.}  We present our ecovisor design, which virtualizes a physical energy system to enable software-based control of server power consumption and battery charging/discharging, as well as visibility into variable grid carbon-intensity and renewable generation. Our ecovisor exposes a software API to applications that enable them to control their use of power to respond to uncontrollable variations in grid energy's carbon-intensity and renewable energy's availability. 

\noindent {\bf Carbon-Efficiency Optimizations.} We present multiple case studies showing how a range of different applications can use the ecovisor API to optimize their carbon-efficiency.  Our case studies highlight two important concepts including:  i) different applications use their virtual energy system in different ways to optimize carbon-efficiency, and ii) application-specific policies can better optimize carbon-efficiency compared to general one-size-fits-all system policies. While optimizing energy-efficiency has been well-studied in computing, there has been little research on optimizing carbon-efficiency, which is fundamentally different and the only metric that really matters for addressing climate change.

\noindent {\bf Implementation and Evaluation.} We implement a small-scale ecovisor prototype on a cluster of microservers that exposes a virtual grid connection, solar array, and batteries to applications. We evaluate our prototype's flexibility by concurrently executing the case study applications above, and showing that optimizing their carbon-efficiency on a shared infrastructure requires application-specific policies.  For example, an interactive web service may use carbon budgeting to maintain a strict latency SLO as carbon-intensity varies, while a parallel batch job might instead adjust its degree of parallelism.

%% file: background.tex
\noindent {\bf Motivation}. Sustainable computing focuses on the design and operation of carbon-efficient computing infrastructure and applications. This paper focuses on reducing Scope 2 operational carbon (and other GHG) emissions from using electricity~\cite{ghg}, which represents a significant fraction of cloud platforms' emissions. Optimizing Scope 1 direct emissions and Scope 3 embodied emissions are outside our scope.  While these other classes of emissions are also important, cloud platforms have few Scope 1 emissions, and have no direct control over their Scope 3 emissions. 

While cloud platforms have long focused on optimizing energy-efficiency, optimizing carbon-efficiency is fundamentally different. To illustrate, consider that a highly energy-efficient system can be highly carbon-inefficient if its grid-supplied power derives from burning fossil fuels, while a highly energy-inefficient system can be highly carbon-efficient if its power derives solely from zero-carbon renewable energy.  As this trivial example shows, a cloud platform's carbon-efficiency depends, in part, on the carbon-intensity of its energy supply, which varies widely over time based on variations in both grid power's carbon-intensity and local renewable power's availability.

Since modifying a cloud platform's operations to adapt to variations in carbon-intensity, e.g., by throttling workloads when carbon-intensity is high, is challenging, cloud providers have largely focused on transparently reducing their \emph{net} carbon emissions using \emph{carbon offsets}. Such offsets are an accounting mechanism that enables offsetting the direct use of carbon-intensive energy by purchasing zero-carbon renewable energy generated at another time and location~\cite{carbon-offsets,google-blog2}. Carbon offsets are attractive because they do not require complex operational changes to reduce net carbon emissions.  Many prominent technology companies have eliminated their net carbon emissions~\cite{amazon-carbon-neutral,google-carbon-free,facebook-carbon-neutral,microsoft-carbon-negative}, which they often refer to as running on ``100\% renewable energy.'' Unfortunately, carbon offsets do not reduce direct carbon emissions, and become increasingly less effective as carbon emissions decrease, as there is less carbon left to offset.  As a result, eliminating \emph{absolute} carbon emissions will ultimately require cloud platforms to change their operations to reduce their direct carbon emissions by better aligning their computing load with when and where low-carbon energy is available.  

Reducing direct carbon emissions is challenging largely because it introduces a new constraint that requires voluntarily making difficult tradeoffs between performance/availability, cost, and carbon emissions.  In general, modifying design and operations to reduce direct carbon emissions decreases performance/availability and also increases cost, as energy prices do not (yet) incorporate the cost of carbon's negative externalities to the environment.   Importantly, the optimal tradeoff between performance/availability, cost, and carbon emissions differs across applications and users.  As we show in \Section\ref{sec:evaluation}, the policies for reducing the carbon emissions of delay-tolerant batch applications are significantly different from those for interactive web services that must maintain a strict latency Service Level Objective (SLO).  More generally, though, cloud users, i.e., companies, have widely different goals, strategies, and tolerances for reducing carbon (at the expense of increased cost and lower performance/availability), which cloud platforms do not know. As a result, cloud platforms are not well-positioned to manage carbon emissions at the system-level on behalf of their users, which motivates our ecovisor's approach of exposing energy and carbon management to applications.

The motivation for our ecovisor's application-level control of carbon is analogous to that for cloud auto-scaling: all cloud platforms support elastic auto-scaling that enables applications to horizontally or vertically scale their resources in response to variations in their workload's intensity \cite{aws-autoscale,azure-autoscale}. These auto-scaling policies are application-specific for similar reasons as above, i.e., differing application requirements and user tradeoffs between cost and performance/availability. Our ecovisor's API, discussed in \Section\ref{sec:design}, enables similar ``auto-scaling'' but in response to variations in grid power's carbon-intensity and local renewable energy's availability.  A simple evolutionary path to enabling such ``carbon-scaling'' using an ecovisor is to augment existing cloud auto-scaling APIs. For example, existing APIs, such as Amazon CloudWatch~\cite{cloud-watch} and Azure Monitor~\cite{azure-monitor}, already expose visibility into platform resource usage, and could easily be extended to include power and carbon information. In this case, cloud platforms would ``delegate'' carbon-scaling to applications just as they currently delegate auto-scaling resources.

While the ecovisor approach could apply to existing cloud platforms, especially those hosted at datacenters with substantial co-located renewables~\cite{maiden} and energy storage~\cite{microsoft}, there is currently no financial incentive to reduce carbon.  This is a social problem, not a technical one.  In the end, to halt climate change, government policies will likely be necessary to create strong incentives for monitoring and reducing carbon emissions, either directly, e.g., via carbon caps, or indirectly, e.g., via carbon pricing.
Nevertheless, cloud platforms have already begun to expose visibility into their carbon emissions~\cite{google-carbon3}, driven by their customers' increasing desire to measure and report carbon emissions data. This combination of customer demand and government policy is likely to incentivize future cloud platforms to adopt ecovisor-like mechanisms for measuring and controlling carbon emissions.

\noindent {\bf Background.} Our work assumes a datacenter's physical energy system connects to up to three distinct power sources:  the electric grid, local batteries (or other forms of energy storage), and local renewable generation, such as solar or wind.   The power supplied to servers (and other computing equipment) is a mix of these three power sources. Not all facilities will have connections to all three power sources, and the capacity of each source may vary.  For example, many large cloud data centers may not have local renewables, while smaller edge sites might not require a grid connection, i.e., if they have enough local renewables and battery capacity to be self-powered~\cite{parasol}.

An ecovisor requires software-defined monitoring and control of both server power and the physical energy system, i.e., power's supply, demand, and carbon emissions.

\emph{Monitoring Power}. An ecovisor must be capable of monitoring each energy source's power generation and consumption.  Energy system components commonly expose power monitoring via programmatic APIs.  For example, battery charge controllers, such as Tesla's Powerwall, support querying a battery's energy level, and its charge/discharge rate from the grid and solar~\cite{tesla-api}, while solar inverters support querying current and historical solar power generation~\cite{tesla-api}.  Our ecovisor builds on these existing APIs.  An ecovisor must also be capable of monitoring server power consumption.  Most servers include power monitoring functions internally, e.g., in hardware exposed to the OS, or externally, e.g., via IPMI~\cite{ipmi}.    

\begin{figure}[t]
\centering
\includegraphics[width=\columnwidth]{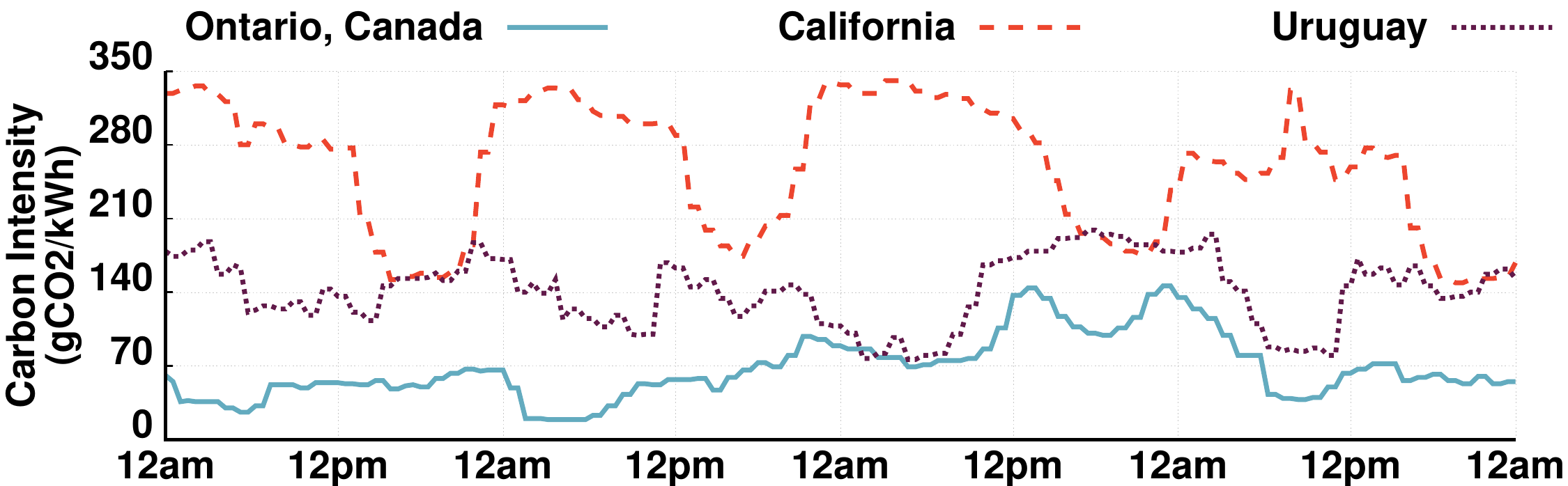}
\caption{\emph{Grid carbon emissions for three different regions showing spatial and temporal variations.}}
\label{fig:carbon_trace}
\end{figure}

\emph{Monitoring Carbon}. An ecovisor must also be capable of monitoring grid power's carbon-intensity in real time.   Recently, third-party carbon information services, such as electricityMap~\cite{electricity-map} and WattTime \cite{watttime}, have begun providing real-time, location-specific estimates of grid power's carbon-intensity by tracking the output of each grid generator, and estimating its carbon emissions based on the type of generator.   Cloud platforms are now using these services to provide coarse-grained estimates of each region's average grid power carbon-intensity~\cite{google-carbon}. Instead, our ecovisor uses these APIs to monitor fine-grained grid carbon-intensity, e.g., every 5 minutes. Figure~\ref{fig:carbon_trace} shows how grid power's carbon-intensity varies over time at different locations.  As shown, Ontario has the lowest carbon-intensity due to its use of nuclear power, while Uruguay has a slightly higher carbon-intensity due to its use of hydroelectricity. California has the highest carbon-intensity due to a higher use of fossil fuels, but also the highest variability due to its high solar penetration. 

\emph{Controlling Power.} Finally, our ecovisor must be capable of controlling power usage in response to changes in grid power's carbon-intensity and renewable energy availability, including regulating server power consumption and battery charging/discharging, i.e., by enabling software to cap the maximum power discharged from batteries and regulate when and how much to charge batteries from the grid and renewables.   In the former case, there has been significant prior work on power capping servers and containers by limiting their resource usage~\cite{google-osdi20,power-containers}.  Our ecovisor leverages these software-based techniques to cap per-container power.  Specifically, our prototype takes a similar approach as recent work~\cite{google-osdi20}, which caps container power by limiting the utilization per core.  In the latter case, battery charge controllers often do not expose control functions to software, since they implement the reliability abstraction, which never artificially caps power and always charges grid-connected batteries to full capacity. However, recently, battery management systems, such as Tesla's Powerwall, have begun to expose these functions in software, which our ecovisor leverages~\cite{tesla-powerwall}.

%% file: design.tex
\begin{figure}[t]
\centering
\includegraphics[width=\columnwidth]{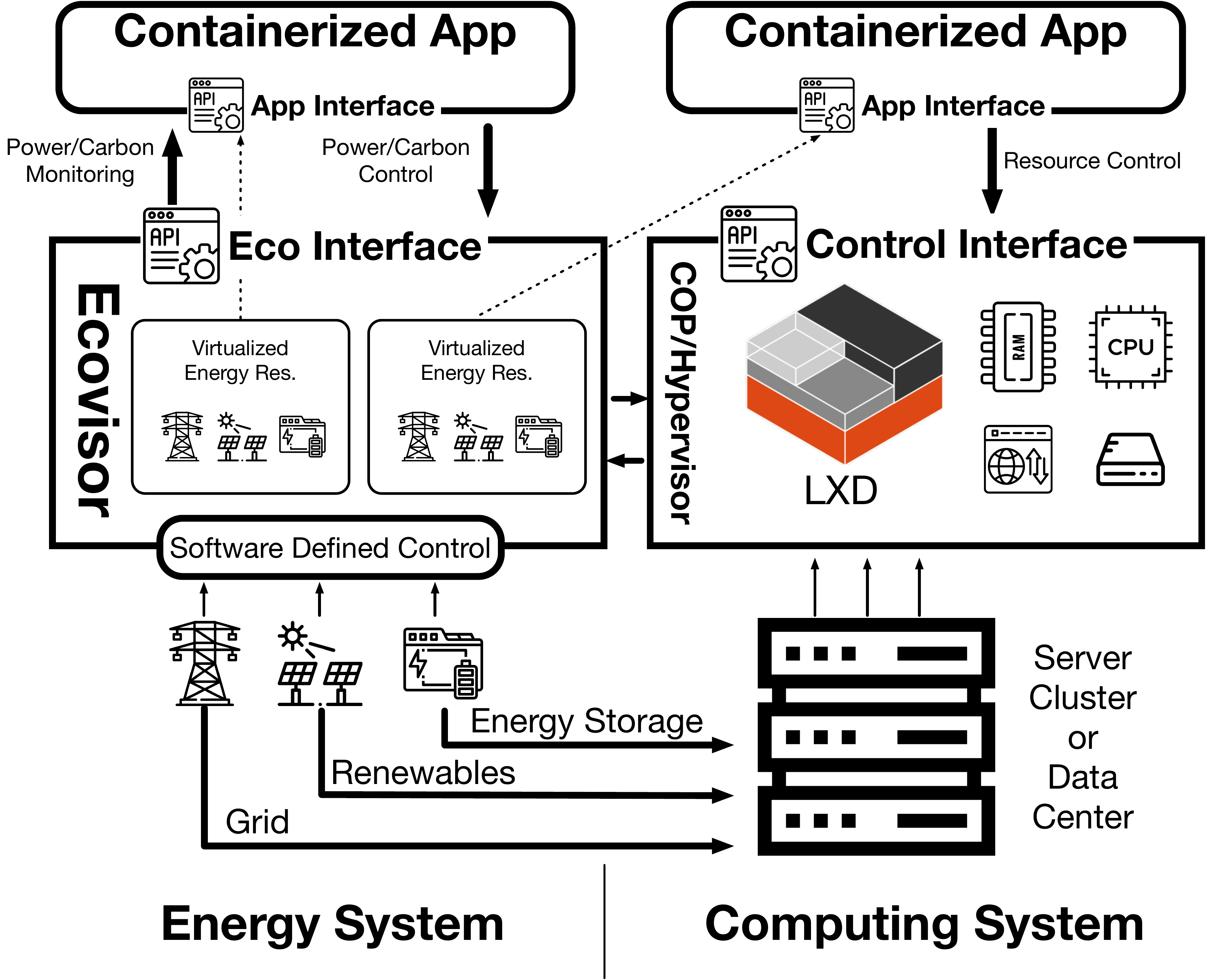}
\caption{\emph{Our ecovisor's design uses containers (or VMs) for resource, energy, and carbon management.}}
\label{fig:system_design}
\end{figure}

Figure~\ref{fig:system_design} provides an overview of our ecovisor's general design, which uses containers or virtual machines (VMs) as the basic unit of resource allocation and energy management. We chose a container/VM instance-level API, in part, because it aligns with, and could easily extend, existing instance-level cloud APIs.  As we discuss, an instance-level API can also support higher-level cluster or cloud-level APIs that provide simplified abstractions for specific applications, such as geo-distributed applications.

An ecovisor integrates with and extends an existing orchestration platform that already provides basic container (or VM) management and monitoring functions, including creating and destroying containers (or VMs), as well as allocating resources to them.  
Note that Container Orchestration Platforms (COPs), such as Kubernetes and Mesos, and similar VM orchestration platforms, generally do not provide sophisticated, fine-grained energy monitoring and management functions.  As discussed in \Section\ref{sec:implementation}, our implementation specifically builds on LXD~\cite{lxd}, which is a simple COP that exposes basic container management functions over a REST API, similar to Kubernetes and Mesos.  We chose to extend a COP for our prototype because these platforms have become the \emph{de facto} operating systems for uniformly managing the resources of large server clusters. However, while we focus our discussion below on COPs, our design also applies to similar platforms that orchestrate VMs
rather than containers. 

\input{table}

COPs provide distributed applications the abstraction of their own virtual cluster composed of multiple containers, each with a specified resource allocation.  These virtual clusters are elastic, such that the number of containers and each container's allocated resources may grow or shrink over time based on application demand and resource availability.  In particular, applications may \emph{horizontally scale} their number of allocated containers as demand changes, or \emph{vertically scale} the resources allocated to each container.  COPs include a scheduling policy that determines how to allocate resources to applications under constraint.  There are many possible resource scheduling policies that optimize for different objectives, such as fairness, e.g., Dominant Resource Fairness~\cite{drf} or revenue, e.g., cloud spot markets~\cite{aws-spot,azure-spot}.   These policies may require the scheduler to reclaim (or revoke) resources from distributed applications.  As a result, distributed applications that run on COPs are already designed to be resilient to resource revocations.  As we discuss, this resiliency is also useful for designing carbon-efficient applications, since the unreliability of low-carbon energy may cause power shortages that also manifest as resource revocations. 

\subsection{Extending COPs with an Ecovisor}
\noindent \textbf{Virtual Energy System.} An ecovisor extends COPs' existing API to provide the abstraction of a virtual energy system, which supplies power to each application's virtual cluster.  As shown in Figure~\ref{fig:system_design}, our virtual energy system includes a virtual grid connection, a virtual battery, and a virtual solar array.   Applications receive a share of grid power, the physical solar array's variable power output, and the physical battery's energy and power capacity.   We focus on solar, instead of wind, because it has higher average power density and is more widely available. 

\noindent \textbf{Ecovisor Interface.} Table~\ref{table:api} shows our ecovisor's narrow API, which is composed of three basic types of methods: getters, setters, and an asynchronous notification.  

\emph{Getter Methods}. The getter and setter methods are synchronous downcalls.   Applications use the getter methods for simple power and carbon monitoring, including retrieving their current virtual solar power output, grid power usage, grid power carbon-intensity, per-container power caps, and per-container power usage.  As discussed in \Section\ref{sec:background}, this information is readily available from the physical energy system's components, servers, and carbon information services.  Our ecovisor provides applications a uniform centralized interface to access this information, and also stores historical data in a time-series database to support sophisticated queries over historical data. 

\emph{Setter Methods}. Applications use the setter methods to control their virtual power's supply and demand.  Applications exercise control over their i) power demand by setting per-container power caps and ii)  power supply by determining when and how fast to charge their battery, as well as when to discharge the battery and the maximum rate of discharge.  The API does not include any functions for controlling virtual solar power, since it is dictated by the environment. Simple datacenters may have only grid power with no renewables or batteries. In this case, applications control their carbon emissions by explicitly setting per-container power caps to regulate grid power in response to variations in its carbon-intensity. Datacenters that also have batteries may also perform carbon arbitrage, e.g., by charging their virtual batteries when carbon-intensity is low and discharging when high, in addition to regulating their grid power usage. 

When solar power is also available, the ecovisor configures an application's virtual energy system to always use virtual solar power first to satisfy demand.  If there is excess solar power after meeting demand, the ecovisor automatically uses it to charge an application's virtual battery.  If an application has configured its virtual battery to charge at a higher rate than the excess solar power, then its virtual energy system supplements the charging up to the specified rate using grid power, and attributes any carbon emissions from using grid power to the application. If an application's virtual battery fills to capacity, its excess virtual solar power must go somewhere:  while resource schedulers can choose whether or not to be work-conserving, physics dictates that our virtualized energy system is energy-conserving.   Determining how to handle excess solar power is a policy decision.  For example, an ecovisor may reclaim excess solar energy and re-distribute it to other applications (if they have available virtual battery capacity), net meter it back to the grid (if possible), or rely on the battery charge controller to curtail it.  

If there is not enough virtual solar power to meet an application's demand, its virtual energy system first uses up to the maximum specified battery discharge rate to satisfy the deficit.  If the maximum specified battery discharge rate is still not sufficient, then the virtual energy system finally uses grid power to make up the difference, and again attributes any carbon emissions from using grid power to the application. Importantly, while grid power's carbon-intensity, solar power, and container power usage vary continuously, our ecovisor discretizes and accounts for these values over a small discrete time (or tick) interval $\Delta t$, e.g., every minute. The virtual energy system always retains a small amount of virtual battery capacity to store the maximum solar power output over the tick interval, and accounts for this solar power output in the next interval.   As a result, applications always know the solar power available to them in the next tick interval. 

\begin{table}[t]
    \footnotesize
    \begin{center}
    	\scalebox{0.95}{
    \begin{tabular}{|| l | l ||} \hline
    {\bf Function Name} & {\bf Description} \\ \hline \hline  
    {\tt get\_container\_energy()} & Energy usage in interval $(t_1,t_2)$ \\ \hline
    {\tt get\_container\_carbon()} & Carbon usage in   interval $(t_1,t_2)$ \\ \hline
    {\tt get\_app\_power()} & Power usage for an application   \\ \hline
    {\tt get\_app\_energy()} & Energy usage in interval $(t_1,t_2)$  \\ \hline
    {\tt get\_app\_carbon()} & Carbon usage for an application   \\ \hline
    {\tt set\_carbon\_rate()} & Set carbon rate for a container \\ \hline
    {\tt set\_carbon\_budget()} & Set application's carbon budget    \\ \hline
    {\tt notify\_solar\_change()}  & Called when solar changes\\  \hline
    {\tt notify\_carbon\_change()}  & Called when grid carbon changes \\ \hline
    {\tt notify\_battery\_full()}   & Called when battery fully charged \\ \hline
     {\tt notify\_battery\_empty()} & Called when battery empty \\ \hline
    \end{tabular}}
    \caption{\emph{Example library functions using ecovisor's API.}}
    \label{tbl:library_functions}
    \end{center}
\end{table}

\noindent {\bf Asynchronous Notifications}.   An ecovisor's virtual energy system abstraction also includes an asynchronous upcall notification based on the tick interval, mentioned above.  The {\tt tick()} method is akin to an OS timer interrupt and triggers at the same tick interval over which the virtual energy system discretizes power.  Applications register their {\tt tick()} method with the ecovisor as a callback function at startup. Within their {\tt tick()} method, applications can examine the characteristics of their power supply, e.g., current solar power output, battery charge level, and grid power's carbon-intensity, and their application's characteristics, e.g., container resource utilization, power usage, and application-level performance metrics, and make adjustments to their power supply and demand to balance potentially competing objectives, such as performance, energy-efficiency, carbon emissions, and cost. 

There are many other external events that might require an immediate application response, which an ecovisor could also expose to applications via asynchronous upcalls.  For example, a significant and sudden change in virtual solar power output or grid power's carbon emissions, or the virtual battery reaching the full or empty state.  However, since we intend the {\tt tick()} method to execute at fine-grained intervals, e.g., every minute, applications are typically able to recognize and address these external events within their {\tt tick()} method.  In general, carbon does not change significantly within a minute, and since our ecovisor always maintains a small amount of battery capacity to buffer solar, the battery never runs empty within a tick interval.   While a virtual battery may fill up within a tick interval, it only has the potential to waste a small amount of excess solar power over the interval. 

Applications may also call container and resource management functions in response to changes in available solar power or grid carbon-intensity.  For example, applications may horizontally scale their number of containers, or the resources allocated to each container, up or down as solar power and grid power's carbon intensity vary.  

\subsection{Library Interfaces}
Our ecovisor's API from Table \ref{table:api} is simple and narrow by design, as it represents the minimal set of functions necessary to control power's supply and demand.  We chose a container-level API to enable the widest range of policies.  Importantly, developers can use these functions to implement a range of higher-level interfaces and abstractions that simplify interactions with the virtual energy system, or make it entirely transparent to applications.  For example, developers could use our container-level API to implement cluster-level policies.  In addition, distributed applications  that control virtual energy systems at multiple sites could implement geo-distributed policies that shift workload to the site(s) with 
the lowest carbon-intensity or most renewable availability.   As a result, the additional complexity of using a virtual energy system need not be borne by most applications, but can instead be encapsulated in third-party software libraries and services, as with exokernels and similar library operating systems.  

An ecovisor promotes innovation by enabling the development of libraries and services that implement a wide range of application-specific energy and carbon management policies. Since users and applications have widely different characteristics, goals, strategies, and tolerances for reducing carbon, which cloud providers do not know, cloud providers are not well-positioned to manage energy and carbon on behalf of their users at the system-level.

Table~\ref{tbl:library_functions} depicts some simple library functions we implemented for \Section\ref{sec:evaluation}'s case studies.  These functions enable applications to monitor their energy usage and carbon emissions over various time intervals, both on a per-container and per-application basis, as well as specify a carbon rate or budget, such that the carbon rate dictates a threshold rate (per unit time) of carbon emissions, while a budget sets a total limit on an application's carbon emissions.  

\subsection{Multiplexing the Physical Energy System}
Each application's virtual energy system exposes an API that is functionally equivalent to the underlying physical energy system. Thus, multiplexing control of the physical energy system among applications' virtual energy systems is straightforward, as it simply requires computing the limit on the maximum battery discharge rates and charging rates across all applications. The ecovisor has privileged access to the physical battery charge controller to set these aggregate limits.  The ecovisor also has privileged access to the container management functions to set per-container power caps by setting limits on resource utilization, e.g., using cgroups.   Finally, the ecovisor has privileged access to the energy and carbon monitoring services of the energy system components, e.g., battery charge controller and solar inverter, servers, and carbon information services, which it uses to perform energy and carbon monitoring and accounting for each application. 

We assume an exogeneous policy determines each application's share of grid power, the physical solar array's variable power output, and the physical battery's energy and power capacity.  For example, public cloud platforms might sell solar and battery shares at each site for some price independently of hardware resources.  While there is also a substantial opportunity for ecovisors to dynamically vary, oversubscribe, or share energy resources among applications, similar to analogous policies for computing resources, such inter-application policies are out of scope.  Our focus is instead on enabling many different intra-application policies for optimizing carbon-efficiency.

%% file: table.tex
\begin{table*}[t]
\footnotesize
\begin{center}
\begin{tabular}{|| l | c | c | c | c ||} \hline
{\bf Function Name} & {\bf Type} & {\bf Input} & {\bf Return Value} & {\bf Description} \\ \hline \hline 
{\tt set\_container\_powercap()} & Setter & ContainerID, kW & N/A & Set a container's power cap \\ \hline
{\tt set\_battery\_charge\_rate()} & Setter & kW & N/A & Set battery charge rate until full \\ \hline
{\tt set\_battery\_max\_discharge()} & Setter & kW & N/A & Set max battery discharge rate \\ \hline \hline \hline
{\tt get\_solar\_power()} & Getter & N/A & kW & Get virtual solar power output \\ \hline
{\tt get\_grid\_power()} & Getter & N/A & kW & Get virtual grid power usage \\ \hline
{\tt get\_grid\_carbon()} & Getter & N/A & g$\cdot$CO$_2$/kW & Get current grid carbon intensity \\ \hline
{\tt get\_battery\_discharge\_rate()} & Getter & N/A & kW & Get current rate of battery discharge \\ \hline
{\tt get\_battery\_charge\_level()} & Getter & N/A & kWh & Get energy stored in virtual battery \\ \hline
{\tt get\_container\_powercap()} & Getter & ContainerID & kW & Get a container's power cap \\ \hline
{\tt get\_container\_power()} & Getter & ContainerID & kW & Get a container's power usage \\ \hline \hline \hline 

{\tt tick()} & Notification & N/A & N/A & Invoked by ecovisor every $\Delta t$  \\ \hline
\end{tabular}
\caption{\emph{Ecovisor's narrow API that provides application's visibility and control over their virtual energy system.}}
\label{table:api}
\end{center}
\vspace{-0.6cm}
\end{table*}

%% file: implementation.tex
We first detail our ecovisor software prototype, and then describe the hardware prototype that it runs on. 

\begin{figure}[t]
\centering
\includegraphics[width=\columnwidth]{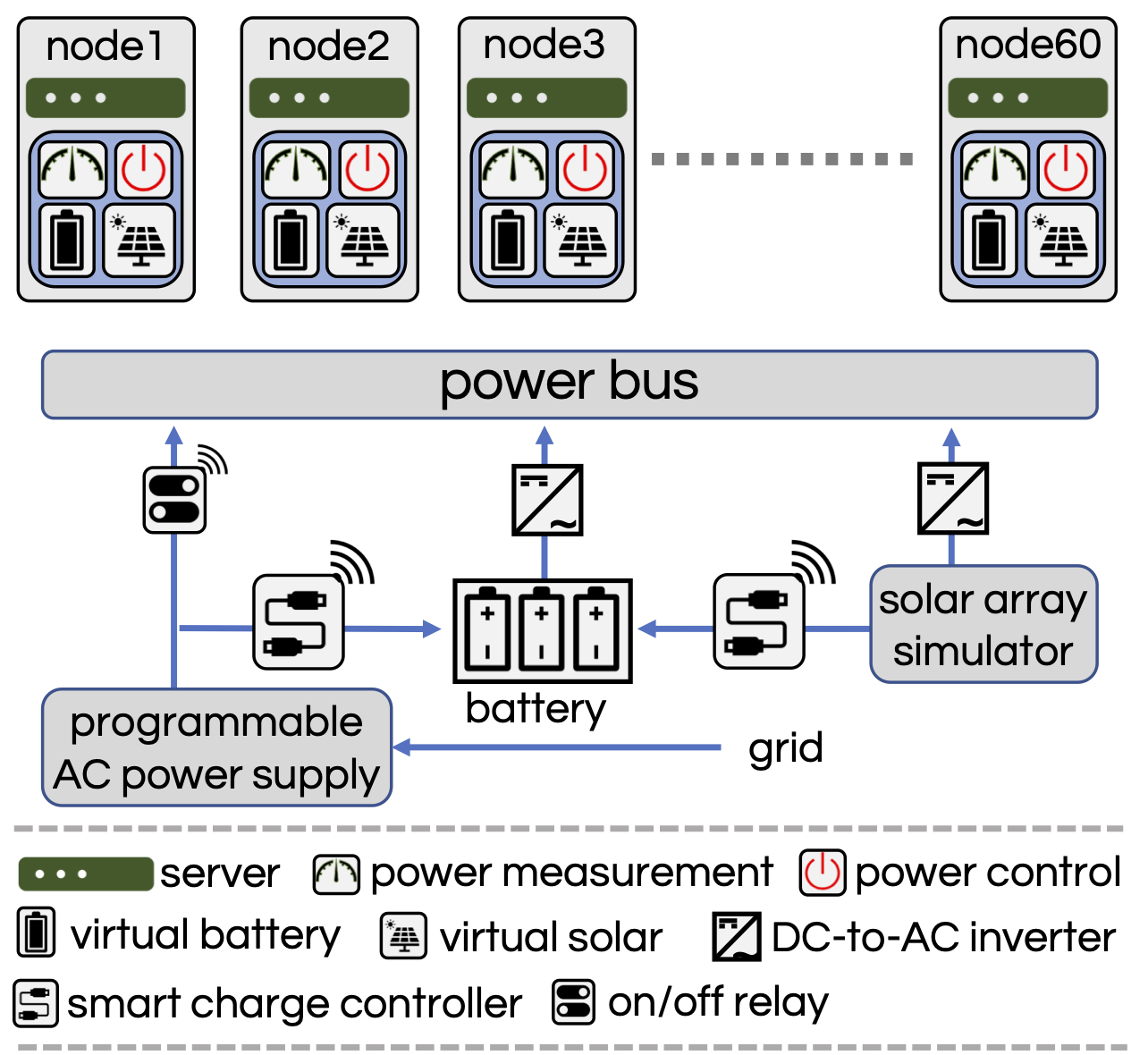}\\
\includegraphics[width=0.9\columnwidth]{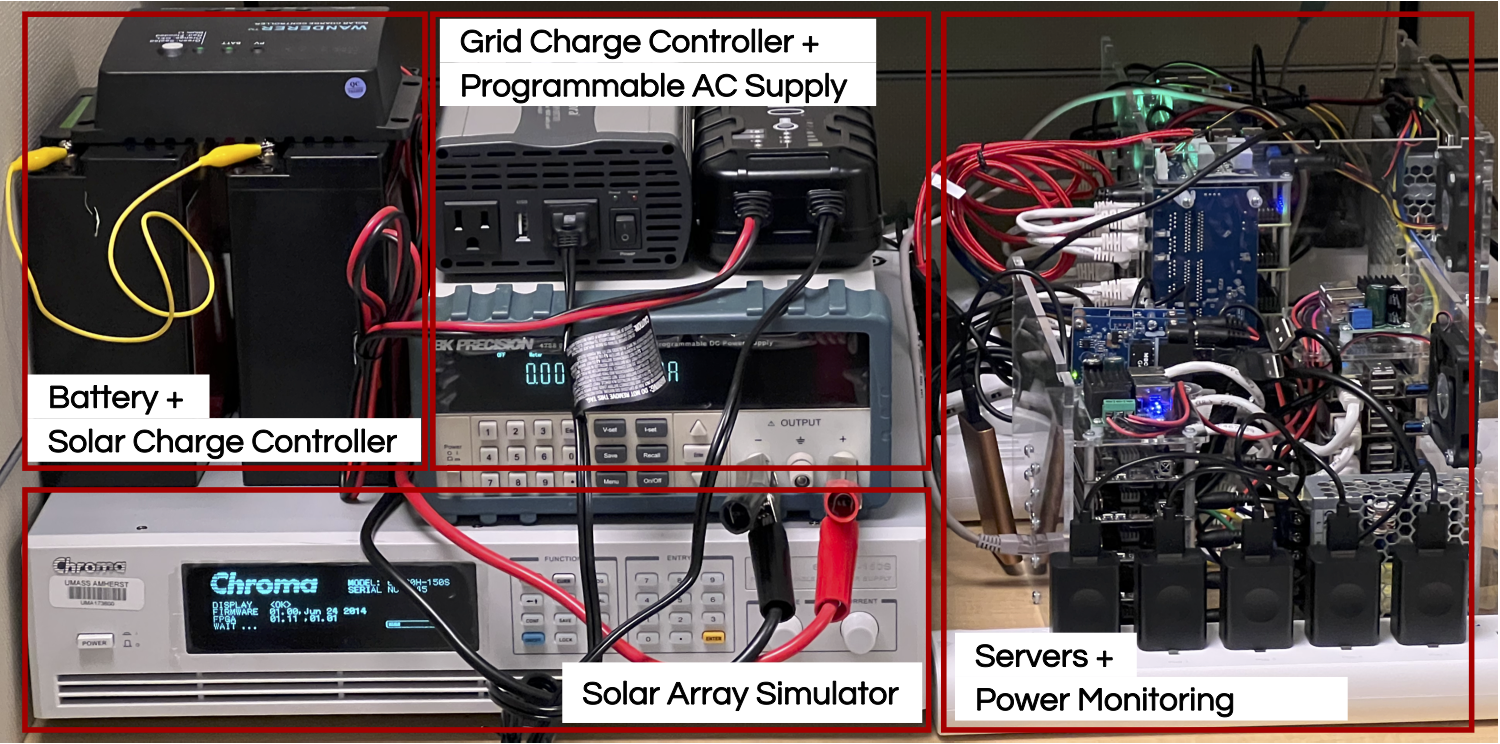}\\
\caption{\emph{Overview of our prototype's physical energy system (top), and a picture of our prototype (bottom).}}
\label{fig:overview}
\end{figure}

\noindent {\bf Software Prototype.} We implemented an ecovisor prototype using Python3 in $\sim$2650 LOC.  Our ecovisor runs on an external server and exposes a REST API to applications that includes the methods from Table~\ref{table:api}. Applications register their {\tt tick()} method as a callback function with the ecovisor server.  Our ecovisor has privileged access to the software APIs exposed by the physical energy system's components and the COP API for monitoring and controlling energy and server resources.  While our approach is generally applicable to any COP, including Kubernetes, our prototype extends LXD~\cite{lxd}, which is a COP that builds on LXC, the Linux container runtime.   We chose LXD due to its flexibility and support for stateful applications and vertical resource scaling. LXD provisions full operating systems within containers (akin to lightweight VMs); enables vertically scaling each container's resources using cgroups; and provides a virtual filesystem (LXCFS) mounted over $/proc$ that provides accurate resource accounting for each container.

Our ecovisor wraps the LXD server, such that applications interact with our ecovisor prototype, which then proxies LXD-specific requests and responses to and from the LXD server.  Our prototype relies on LXD for container management, including horizontal and vertical scaling. Our prototype uses LXD functions internally to vertically scale each container's maximum resource allocation using cgroups to enforce per-container power caps set by the application, as in recent work~\cite{google-osdi20}.   We use PowerAPI~\cite{power-api}, a middleware toolkit for building software-defined power meters, for monitoring power, including per-container power usage, battery power usage, solar power generation, grid power usage, and grid carbon-intensity.  PowerAPI stores this historical power data in a time-series database, specifically InfluxDB, which enables queries over different time intervals.  We use LXD's default container scheduler, which simply allocates a container to the server with the fewest container instances.   

We use electricityMap's API to get grid power's carbon-intensity in real-time. The implementation of other functions in our ecovisor API from Table~\ref{table:api} are hardware-specific. We discuss our hardware prototype below. 

\noindent {\bf Hardware Prototype.}  We built a small-scale hardware prototype of a software-defined physical energy system as a proof of concept.  Figure~\ref{fig:overview} provides an overview and picture of our hardware prototype, which is a cluster of ARM-based microservers, some of which have an attached NVIDIA Jetson Nano GPU. In particular, each microserver includes a quad-core ARM Cortex A53 64-Bit processor and 4GB 1600MHz LPDDR3 memory~\cite{rock64-board} powered by a 2A, 5V power supply. The microservers consume 1.35W at idle, 5W at 100\% CPU utilization, and 10W at 100\% CPU and GPU utilization.  The microservers run Ubuntu 18.04 Bionic minimal 64bit (arm64) with Linux kernel 4.4.2. The smart USB hubs plug into our power bus, which connects to our three power sources---the grid, battery, and solar power---discussed below.  We also implemented simulated versions of each power source to enable experimentation on a more conventional data center server cluster composed of 16 Dell PowerEdge R430s with Intel Xeon processors with 16 cores and 64GB memory.  Implementing a real prototype at this scale is infeasible due to our lab's power constraints, component availability, and cost.  For example, the Chroma 62020H-150S, discussed below, used in our microserver prototype costs nearly \$10,000 and is only capable of emulating a solar array up to 2kW DC. 

\emph{Grid Power.}  To validate the efficacy of our software-based power caps, we connected our system to a programmable power supply that was capable of accurately monitoring grid power consumption.  We used this capability to verify that our system's power usage never exceeded the limit dictated by the container power caps. 

\emph{Battery Power.}  Our prototype's battery bank included multiple 12V, 20Ah deep discharge lithium-ion batteries with a total of 1440Wh capacity.  We configured our battery charge controller to only discharge them to 70\% depth, such that we classify a 30\% state-of-charge as ``empty,'' since deep discharges significantly reduce a battery's cycle life. Our battery can support operating the cluster at maximum power for one hour.  We set the maximum charging rate for the battery bank to 0.25C, which corresponds to 30 amps (A) at 12V, such that the battery charges to full capacity in 4 hours.  We set the maximum discharge rate to 1C, or the rate required to fully discharge the battery in 1 hour.  This rate corresponds to 1440W, which is well above the cluster's maximum power. 

The battery above connects to two smart charge controllers, which expose software APIs: one connected to the grid and the other to solar.  Our ecovisor can use the grid-connected charge controller to set the battery's charging rate. The solar-connected charge controller automatically uses any excess solar power to charge the battery.  Since our prototype does not net meter solar power, we set the charge controller to curtail any excess solar power once the battery is fully charged. 

\emph{Solar Power.}  Our prototype uses a Solar Array Emulator (SAE) instead of a real solar array to enable repeatable experiments. Our SAE is capable of replaying solar radiation traces, and acts like a programmable power supply that mimics the electrical response of a solar module's IV curve.   Thus, we can replace our SAE with a real solar array without requiring any changes. As mentioned above, we use the Chroma 62020H-150S as our SAE, which is widely used for testing solar modules in industry.

%% file: evaluation.new.tex
\begin{figure}[t]
\centering
\includegraphics[width=\columnwidth]{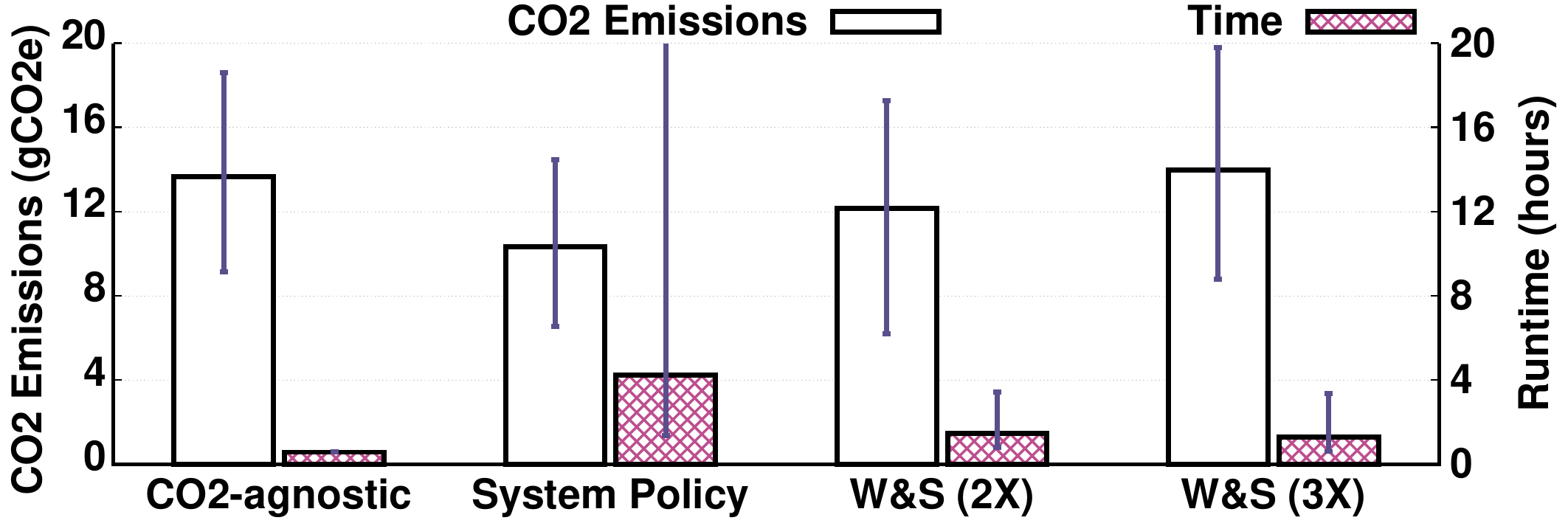}\\
(a) PyTorch ML Training\\
\includegraphics[width=\columnwidth]{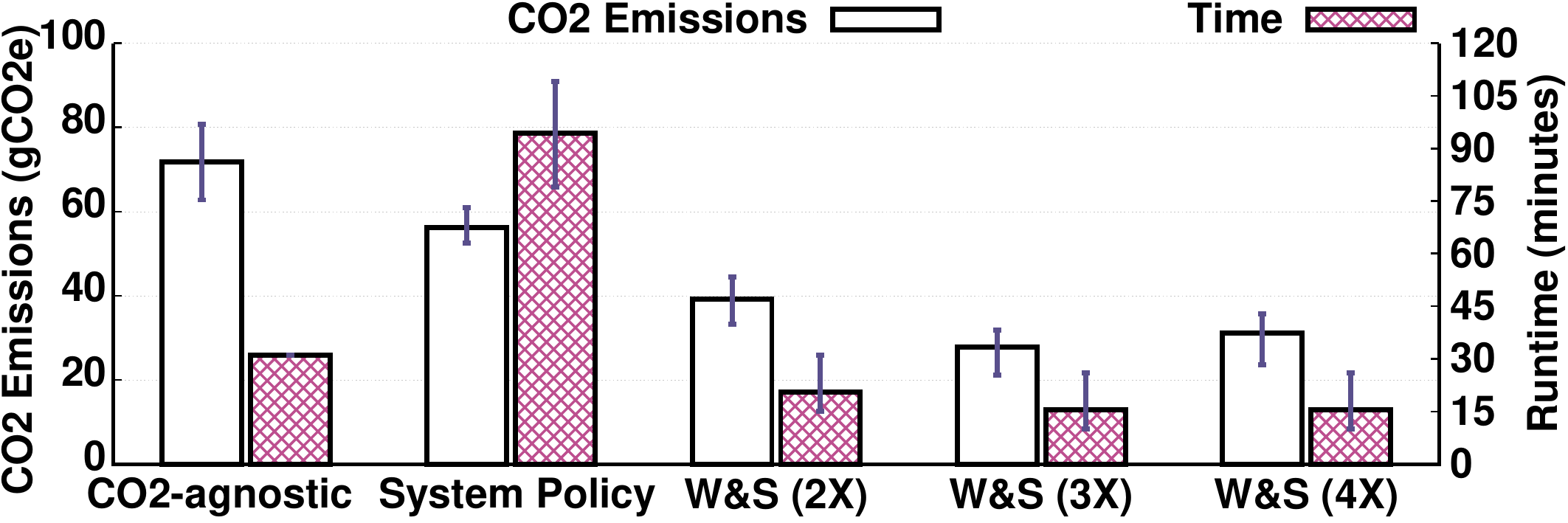}\\
(b) BLAST \\
\caption{\emph{Carbon emissions and runtime for distributed ML training and a high performance computing application using different carbon reduction policies.}} 
\label{fig:reducing-carbon}
\end{figure}

The purpose of our evaluation is to highlight the rich policy space defined by our ecovisor's narrow API and demonstrate that optimizing carbon-efficiency on a shared infrastructure requires application-specific policies. Specifically, we show how our ecovisor can enable a range of different applications to better optimize their carbon-efficiency using an application-specific policy compared to a general one-size-fits-all system policy. Importantly, these applications can operate concurrently on the same infrastructure.  In some cases, we re-implement and improve upon applications from prior work implemented on dedicated platforms~\cite{wait-awhile}.  Of course, our evaluation does not cover all possible uses of an ecovisor, as we expect there are numerous potential carbon-efficiency optimizations and  abstractions for different types of applications that have yet to be developed.  A key goal of our system is to enable the development of these new optimizations and abstractions, while also supporting existing policies.

\subsection{Reducing Carbon}
\label{sec:reducing}

A simple approach to optimizing carbon-efficiency is to suspend execution when grid power's carbon-intensity increases beyond some threshold and resume it later when the carbon-intensity falls below this threshold. Recent work, called WaitAWhile, quantifies the tradeoff between carbon emissions and job completion time using this approach~\cite{wait-awhile}. WaitAWhile's \emph{suspend-resume} policy is an example of a general system-level policy that can be applied to all applications on a shared platform. We compare this suspend-resume policy to a new Wait\&Scale (W\&S) policy we developed, which suspends execution above a threshold and opportunistically scales up resource (and energy) usage when carbon emissions are below the threshold. Wait\&Scale is an application-specific policy, as different applications have different optimal scale-up factors, which the system may not know. Thus, applications are better positioned to configure their optimal scale-up factor based on their specific scaling properties.

\begin{figure}[t]
    \centering
    \begin{tabular}{c}
    \includegraphics[width=0.97\columnwidth]{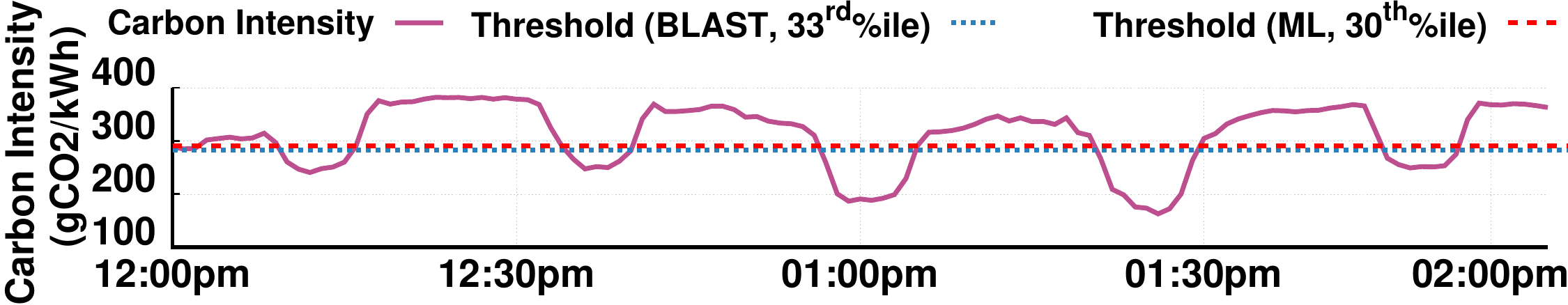}\\
    (a) Carbon intensity and resume thresholds\\
    \includegraphics[width=0.97\columnwidth]{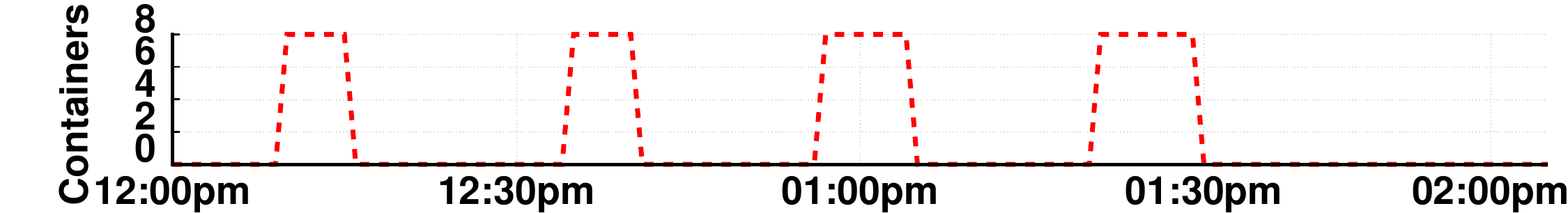}\\
    (b) ML Training with W\&S (2X)\\
    \includegraphics[width=0.97\columnwidth]{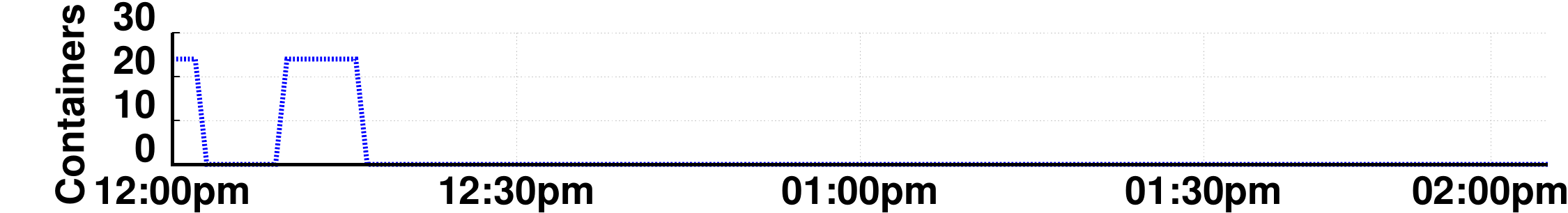}\\
    (c) BLAST with W\&S (3X)\\
    \includegraphics[width=0.97\columnwidth]{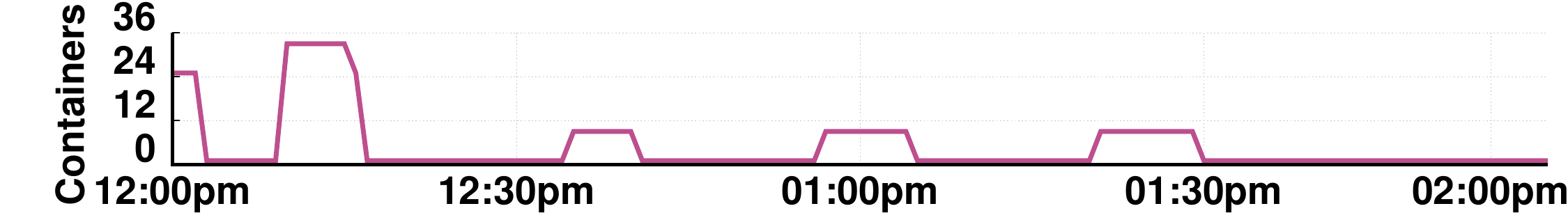}\\
    (d) Ecovisor cluster\\
    \end{tabular}
    \caption{\emph{Multi-tenancy of application-specific carbon reduction policies that use the system resources differently based on their scaling properties.}}
    \label{fig:reducing-carbon-multi}
\end{figure}

\subsubsection{Applications.} Our first experiment runs two applications on a shared multi-tenant infrastructure with different scaling behavior, which characterize each application's speedup as its number of workers increase. 

Our first application is PyTorch, a popular machine learning framework that we use to
train a Resnet34 model~\cite{resnet34} on the CIFAR100 dataset~\cite{cifar100} for five epochs.  The model training job runs on grid power with a variable carbon footprint, which we simulate using data from the carbon emissions of the California Independent System Operation (CAISO)~\cite{caiso} in 2020.  Since carbon emissions vary, we ran the experiment ten times and randomly selected the job arrival each time.  We set the carbon threshold based on the 30$^{th}$\%ile of carbon-intensity over a 48 hour window in each run.  

Our system-level suspend-resume and carbon-agnostic policies run the job on 4 cores, while we run Wait\&Scale with scale factors of 2$\times$ and  3$\times$, which scale up the job to 8 and 12 cores, respectively, when below the carbon threshold. 

The second application is NCBI-BLAST (Basic Local Alignment Search Tool), which is a popular parallel  application that searches for similarities in nucleotide or protein sequences~\cite{blast}.  We use an elastic version of BLAST-470, which can horizontally scale the number of containers it uses at runtime~\cite{tigres}. Our system-level and carbon-agnostic policies run the  BLAST job on 8 cores, while we run Wait\&Scale with scale factors of 2$\times$, 3$\times$, and 4$\times$, on 16, 24, and 32 cores, respectively. We set the carbon threshold based on the 33$^{rd}$\%ile of carbon-intensity over the trace duration.

\subsubsection{Comparing Carbon Reduction Policies.}

Figure~\ref{fig:reducing-carbon} shows the completion times and carbon emissions under different policies 
for the two applications, where the error bars depict the standard deviation across the ten runs.
In both cases, the carbon-agnostic policy has the lowest completion time at the cost of higher carbon emissions. The system-level suspend-resume policy reduces carbon emissions by $24.5$\% and $25.01$\%, 
but frequent suspensions increase the running time by $7.4\times$ and $5.1\times$ for the ML training (top) and BLAST (bottom) applications, respectively.  The system-level policy also exhibits a highly volatile job runtime, since jobs that happen to start executing during a long high-carbon period are forced to stop and wait until the carbon-intensity decreases. 

Wait\&Scale overcomes the high completion times of suspend-resume by opportunistically scaling up resources upon resumption.  For the ML training application (top), Wait\&Scale (2$\times$) achieves a comparable carbon reduction to suspend-resume, but with a significantly lower runtime penalty (of $2.58\times$). However, further scale up does not provide additional carbon benefits---Wait\&Scale (3$\times$) increases carbon emissions by $14.94$\% (similar to the system-level policy) while reducing the runtime by only $12.3$\%.   In this case, scaling up requires more coordination among nodes, which causes synchronization delays that limit speed-up and decrease energy-efficiency.

Unlike Resnet training, BLAST is embarrassingly parallel, and thus scales up much more efficiently when carbon intensity decreases. Wait\&Scale (2$\times$) achieves a carbon reduction of $30.1$\%,  while also reducing runtime by $78.15$\% compared to the system-level policy. Scaling up even further is also beneficial as Wait\&Scale (3$\times$) decreases the carbon emissions by $50.05$\% compared to system-level policy, while further reducing runtime by $83.4$\%.  
The benefits of scaling eventually diminish at 4$\times$ where carbon emissions start increasing, but the job runtime remains the same. For BLAST-470, this happens because BLAST's central queue server becomes a bottleneck when serving tasks to more than 3$\times$ workers.

Importantly, our experiments show that our application-specific Wait\&Scale policy outperforms the system-level suspend/resume policy. In this case, our ML training job and BLAST exhibit different synchronization overheads, which necessitates using different application-specific scale-up factors for optimizing carbon-efficiency. 

\noindent\emph{\textbf{Key Takeaway.} An ecovisor enables applications to optimally configure their scale-up factor to better optimize carbon-efficiency compared to a system-level suspend-resume policy, which is application-agnostic.}

\subsubsection{Multi-tenancy.}
Our experiment above concurrently ran the ML training job and BLAST on a shared multi-tenant infrastructure. Figure~\ref{fig:reducing-carbon-multi} shows the per-application and system-wide power usage for both applications at their optimal scaling factor. Each application uses a different amount resources and power based on their scaling behavior to individually optimize carbon-efficiency. Note that the system-wide power also shows a small amount baseline power required to run the ecovisor.

\begin{figure}[t]
    \centering
    \begin{tabular}{c}
    \includegraphics[width=0.96\columnwidth]{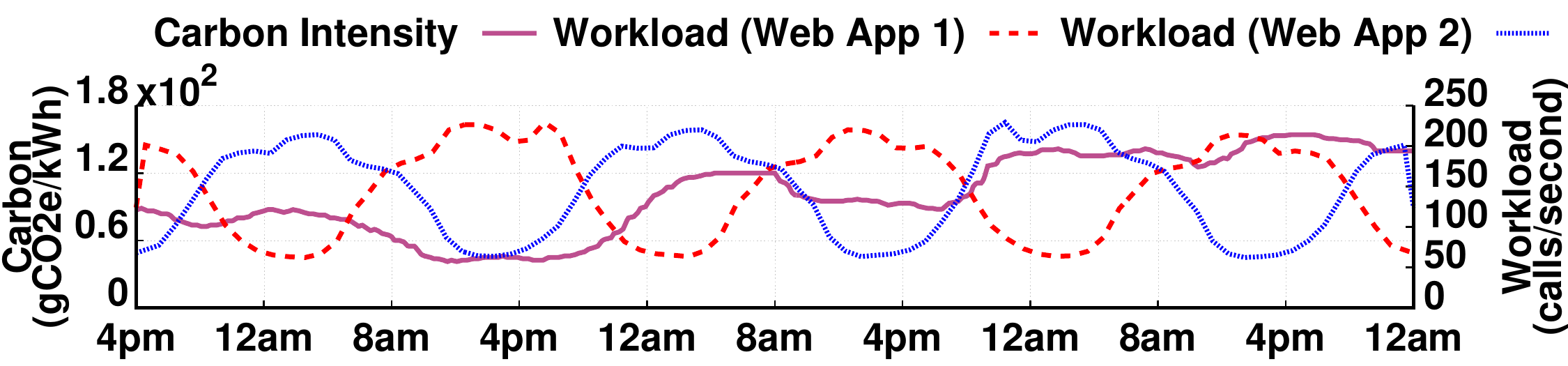}\\
    (a) Carbon intensity and workload\\
    \includegraphics[width=0.97\columnwidth]{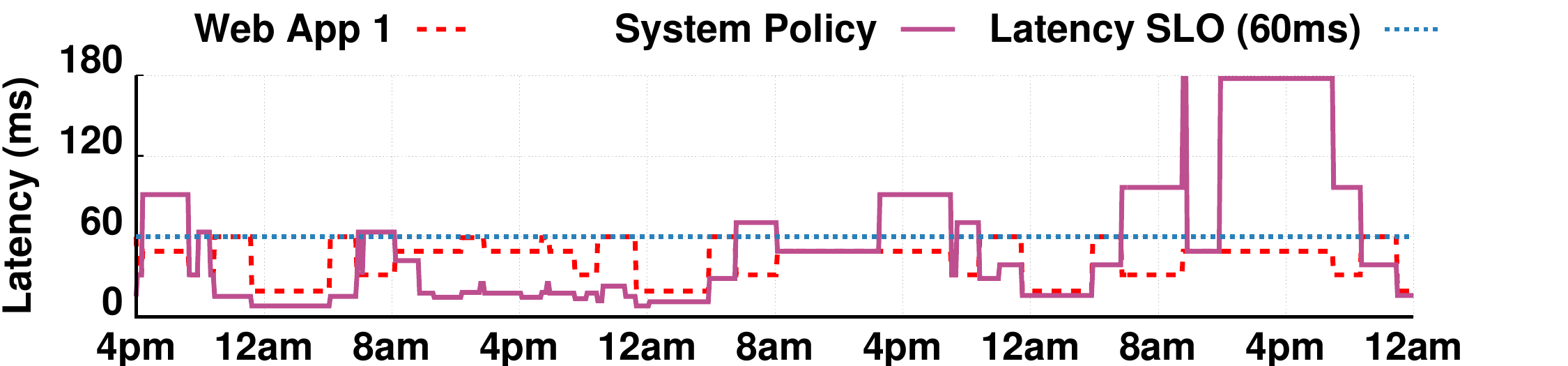}\\
    (b) 95$^{th}$\%ile latency (Web App 1)\\
    \includegraphics[width=0.97\columnwidth]{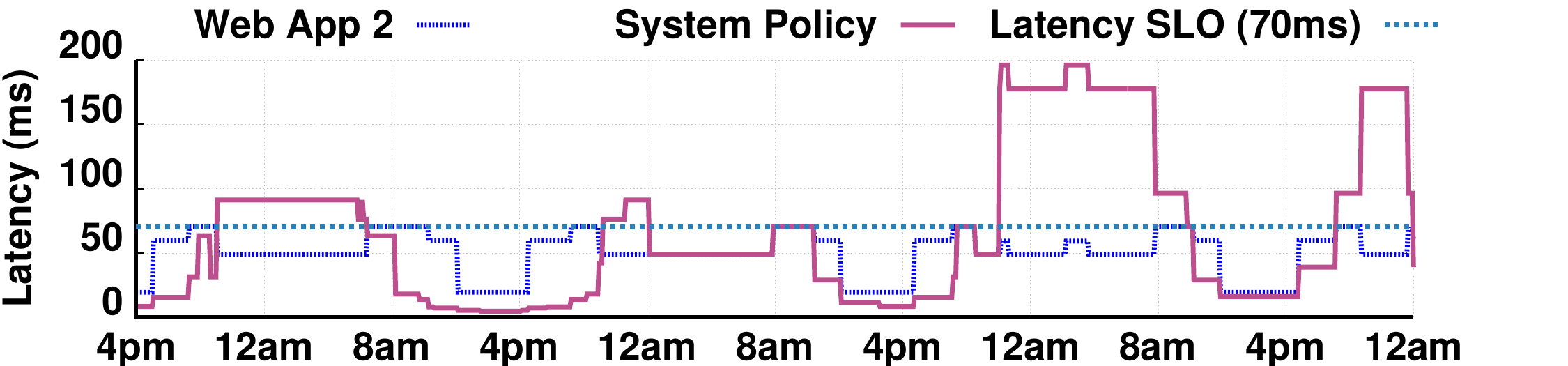}\\
    (c) 95$^{th}$\%ile latency (Web App 2)\\
    \end{tabular}
    \caption{\emph{Comparing a system-level carbon rate-limiting policy with two application-specific carbon budgeting policies for a distributed web application ((b) and (c)) under varying workloads and carbon intensity (a).}}
    \label{fig:webserver-app}
\end{figure}

\begin{figure}[t]
    \centering
    \begin{tabular}{c}
    \includegraphics[width=\columnwidth]{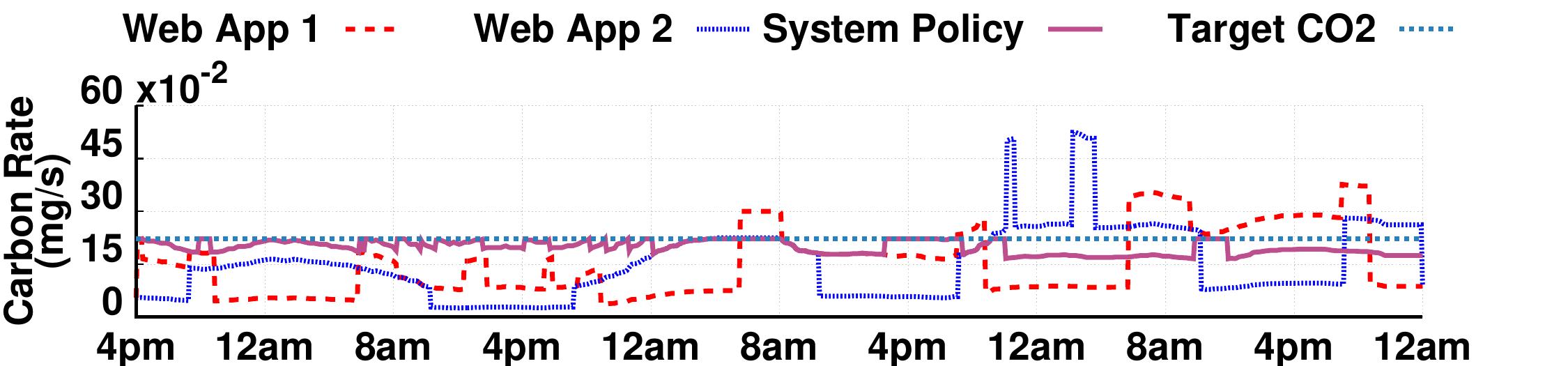}\\ 
    (a) Carbon rate\\
    \includegraphics[width=\columnwidth]{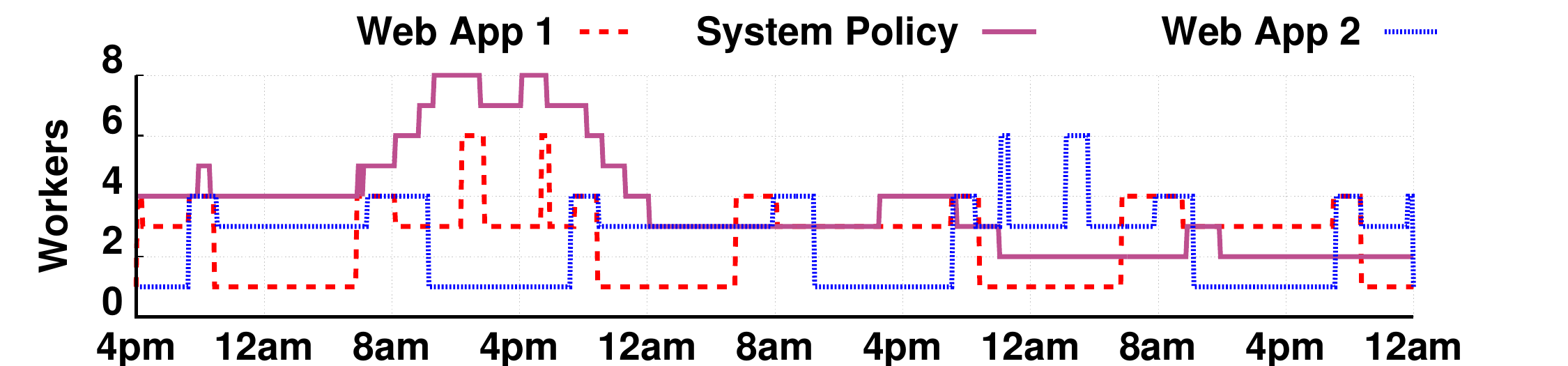}\\
    (b) Workers\\
    \end{tabular}
    \caption{\emph{Multi-tenancy of application-specific carbon budgeting policies that use their system and energy resources differently based on their workload.}}
    \label{fig:webserver-app-multi}
\end{figure}

\subsection{Budgeting Carbon}
\label{sec:budgeting_carbon}
A disadvantage of the above suspend-resume-style carbon reduction policies is that applications cannot make any forward progress during high carbon periods. Furthermore, the goal of reducing carbon emissions may not be suitable for all the applications, which may instead have a specific carbon emissions budget to operate within. For such scenarios, we consider  an application-agnostic system-level policy that enforces a static carbon budget for each application by rate-limiting (or carbon-capping) it at all times.  
We compare this policy to application-specific policies that enforce a more flexible carbon budget over longer time windows, rather than at all times, which allows applications to breach the cap for short periods, if necessary. We show that such dynamic budgeting policies can provide better performance during periods when both the carbon intensity and workload intensity are high.

\subsubsection{Applications.} 
To illustrate the benefits of application-specific dynamic carbon budgeting, we deploy two multi-tenant distributed web applications using our ecovisor prototype. Both applications include a front-end load balancer that distributes web requests across a cluster, and serves a copy of Wikipedia.  The applications use horizontal scaling to regulate power by adding and removing containers from the load balancer's active set.  We subject the applications to two different variable workload demand patterns based on a real-world trace covering 48 hours, and record the latency to satisfy requests~\cite{wikipedia-trace}.  We first run the applications using a static carbon rate limit of  20 mg$\cdot$CO$_2$ per second, and then under a dynamic carbon budget equivalent to the product of the same rate and the trace's length.  Our dynamic carbon budgeting policy horizontally scales the containers up and down to enforce an SLO on the 95$^{th}$\% latency of 60ms and 70ms for the first and second web application, respectively.

\subsubsection{Comparing Carbon Budgeting Policies.}

Figure~\ref{fig:webserver-app}(a) shows the variations over time for the carbon-intensity and workload patterns, which are not aligned.  That is, there are periods of both high carbon and workload intensity. Figure~\ref{fig:webserver-app}(b) and (c) then shows the 95$^{th}$\% response time latency for the two web applications over time. As shown in both (b) and (c), the system-level policy violates the latency SLO near the end of the trace during a period of both high carbon and workload intensity, since it does not have the flexibility to increase its container capacity beyond the static carbon cap to handle the more intense workload. In contrast, the dynamic budgeting policy always satisfies the latency SLO over the entire trace by using fewer resources (and less carbon) during periods of low workload and carbon intensity.  The policy then uses its accumulated ``carbon credits'' to temporarily exceed the carbon rate to serve more intense workloads during high carbon intensity periods, while enforcing the overall carbon budget over a longer time period.

Note that the system-level rate-limiting policy occasionally provides lower latency than the SLO (by over-provisioning when carbon is low), while the dynamic budgeting policy uses fewer resources when they are not needed and leverages the carbon savings to satisfy load spikes. 
Overall, the dynamic  budgeting policy has 22.8\% and 23.4\% lower carbon emissions for both applications compared to the system-level policy, since it operates well below the target carbon rate most of the time. This also demonstrates that the application-specific policy enables the applications to dynamically manage their emissions in different ways while satisfying the overall budget, which is not possible using the system-level rate-limiting policy.

\begin{figure}[t]
    \centering
    \begin{tabular}{c}
    \includegraphics[width=0.97\columnwidth]{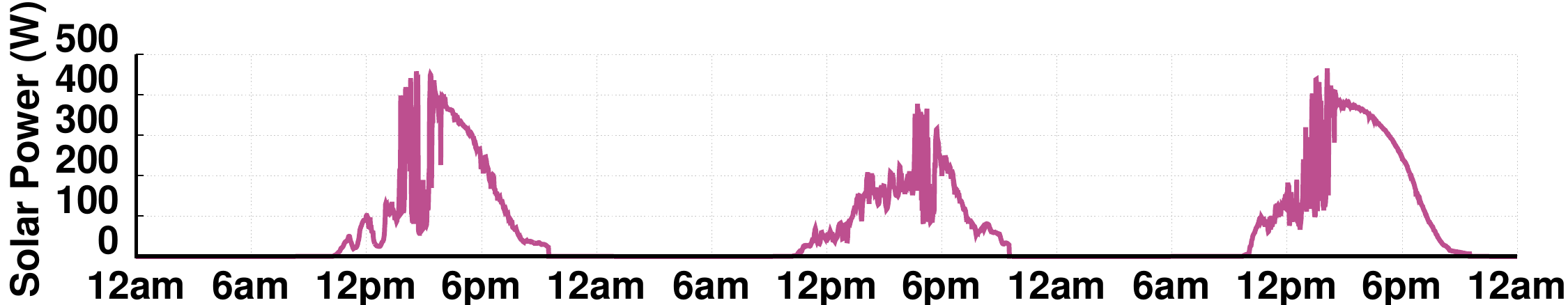}\\
    (a) Solar power output\\
    \includegraphics[width=0.97\columnwidth]{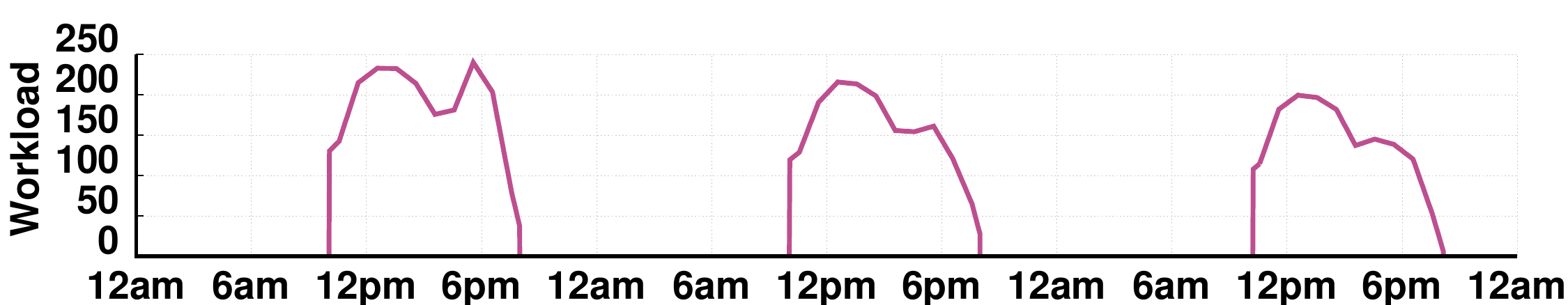}\\
    (b) Web app workload\\
    \includegraphics[width=0.97\columnwidth]{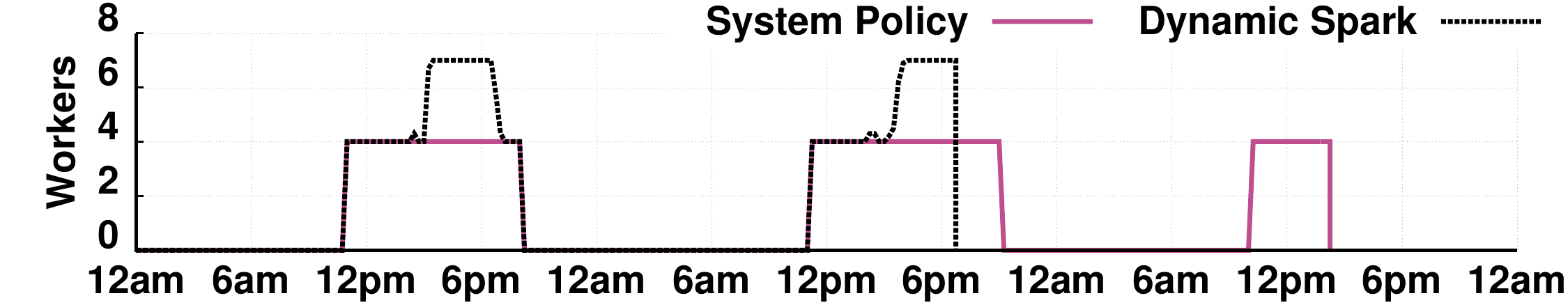}\\
    (c) Workers for static and dynamic Spark\\
    \includegraphics[width=0.97\columnwidth]{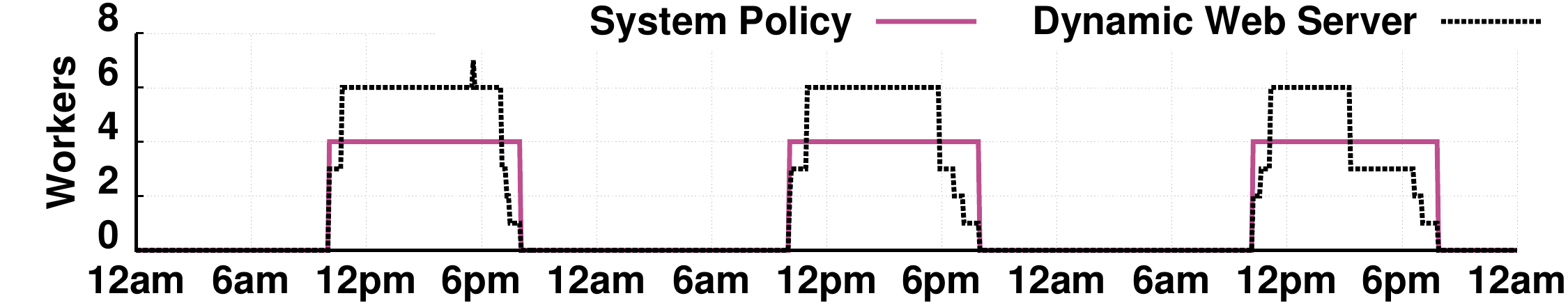}\\
    (d) Workers for static and dynamic web app\\
    \includegraphics[width=0.97\columnwidth]{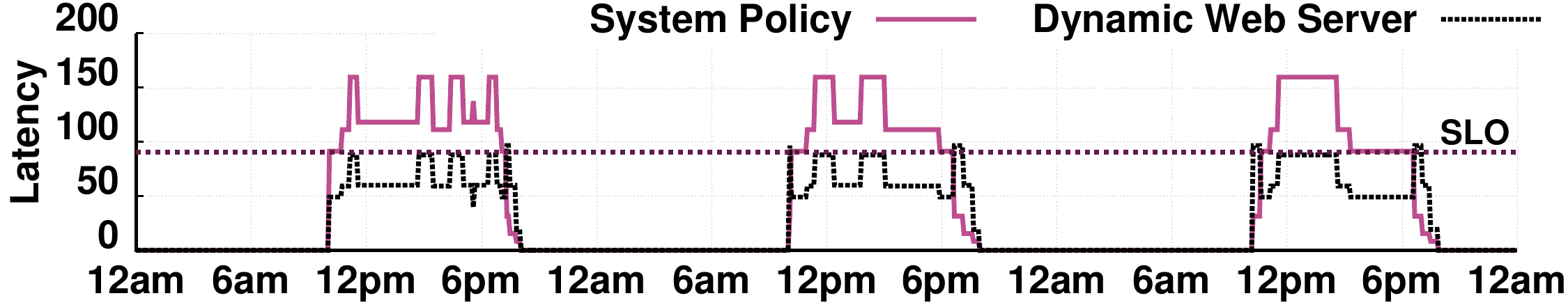}\\
    (e) 95$^{th}$\%ile latency\\
    \end{tabular}
    \caption{\emph{Cluster-level solar power (a) equally divided between Spark and web application, and workload for web application (b). No. of workers with static system and Spark-specific dynamic battery usage policies (c). No. of workers (d) and 95$^{th}$\%ile latency (e) with static system and web app-specific dynamic battery usage policies.}}
    \label{fig:leveraging-batteries}
\end{figure}

\noindent\emph{\textbf{Key Takeaway.}} Applications can better manage a given carbon budget to meet their performance requirements compared to a static system-level rate-limiting policy.

\subsubsection{Multi-tenancy.}
Finally, Figures~\ref{fig:webserver-app-multi}(a) and \ref{fig:webserver-app-multi}(b) show the carbon rate and the number of containers for the two applications over time. 
Both applications, when using dynamic budgeting policy, consume fewer resources and energy when carbon emissions are low, i.e., only enough to satisfy their SLO, while the system-level policy uses as many resources and energy to satisfy its target carbon rate. Although the applications run on the same cluster at the same time, their carbon emissions and container capacity differ depending on their workload.

\subsection{Leveraging Virtual Batteries} 
The applications above optimize carbon-efficiency using grid power. We next examine applications that
implement zero-carbon policies using solar power and batteries. Although solar power has zero carbon intensity,
its output is volatile due to changing environmental conditions. Our ecovisor's virtual batteries can supply applications a minimum guaranteed amount of power when solar output falls below a threshold, smoothing out the volatility.

\begin{figure}[t]
    \centering
    \begin{tabular}{c}
    \includegraphics[width=0.97\columnwidth]{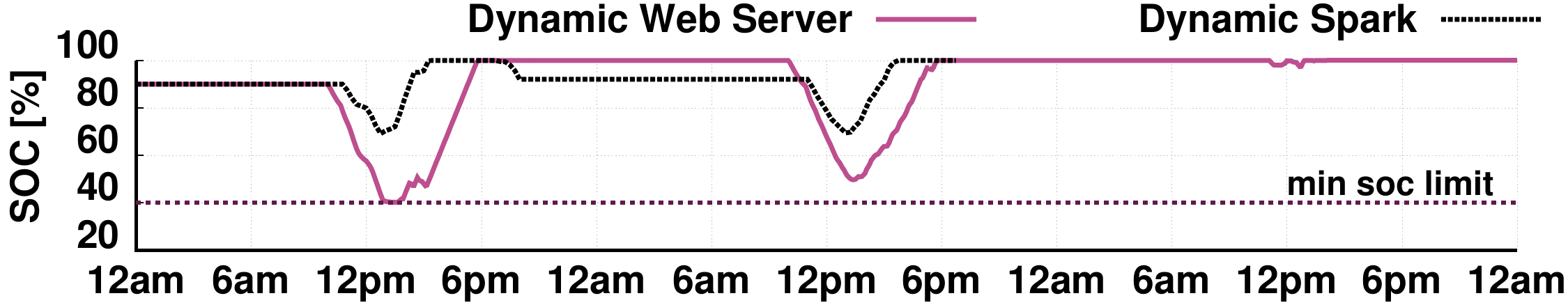}\\
    (a) State of charge for virtual batteries\\
    \includegraphics[width=0.97\columnwidth]{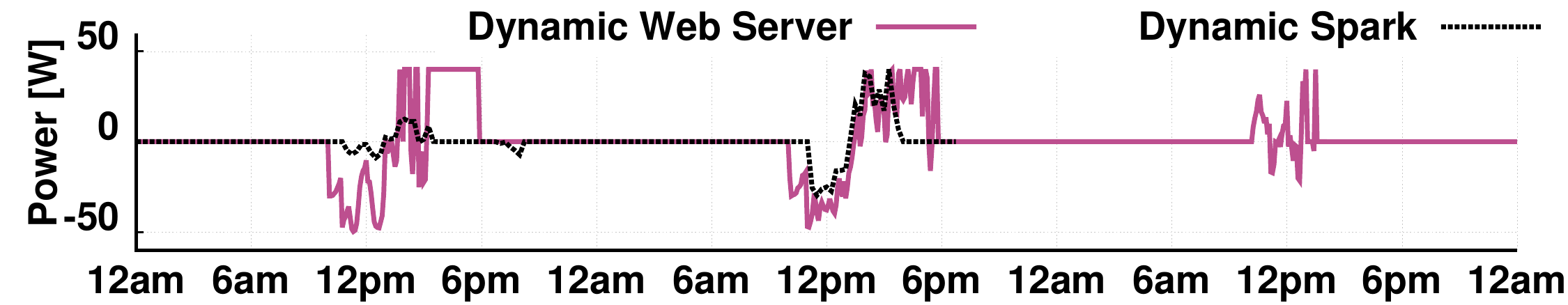}\\
    (b) Virtual battery charging/discharging rates\
    \end{tabular}
    \caption{\emph{Multi-tenancy of application-specific virtual battery usage policies, where each application uses their virtual battery differently based on their requirements.}}
    \label{fig:batteries_multi}
\end{figure}

\subsubsection{Applications.}

To illustrate our zero-carbon policies, we deploy two applications that share a solar panel and physical battery.
Our first application is a ``delay-tolerant'' distributed Spark job running on our ecovisor prototype  powered by intermittent solar and a battery.  In this case, the job is an image preprocessing and feature extraction task written using pyspark running on Spark 3.2.0. Spark runs on solar power and a battery during the day, with the battery ensuring a minimum guaranteed power. Although grid power is available at night, to maintain a zero carbon footprint, we checkpoint completed operations in the Hadoop Distributed File System (HDFS), and wait until the next morning to resume Spark computations. Incomplete workers are terminated without checkpointing every evening and their in-memory results are lost.  

Our second application is a web-based monitoring and logging application, which monitors and logs the power generation of our ecovisor's physical and virtual solar arrays. This application is similar to publicly-deployed services for monitoring renewable-powered computing infrastructure~\cite{solarprotocol,lowtechmagazine}. Each web request logs the current power generation in the web application. Since there is no solar generation at night, the application sees only a daytime workload and is dormant during nighttime hours when there is no data to log. Like Spark, the web application runs on solar power and  batteries during the day and stays suspended during the night. We set a target latency SLO of 100ms for web request processing.

\subsubsection{Comparing Battery Usage Policies.}
Our system-level policy for both applications is to use the battery to smooth out the variations in solar power and provide a minimum guaranteed power. 
Figure~\ref{fig:leveraging-batteries}(a) shows that the total solar power is equally divided between the two applications. 
Figure~\ref{fig:leveraging-batteries}(c) shows the number of workers for a static (system-level) and Spark-specific dynamic policies. 
The system-level policy is conservative and avoids losing computation by using a fixed number of workers that are always available. 
In contrast, the Spark-specific dynamic policy opportunistically scales up the number of workers to leverage excess solar when the battery is fully charged. While any work performed by the additional workers might be lost if they are killed before checkpointing the work, they mostly perform useful computation, which reduces the application runtime by 39\%. 

Figure~\ref{fig:leveraging-batteries}(b) shows the workload trace for the web application, which varies over time as the number of applications running on our ecovisor prototype and monitoring/logging their resources come and go.  Figure~\ref{fig:leveraging-batteries}(d) shows the number of workers for a static (system-level) and application-specific dynamic policies. Since the static (system-level) policy only has fixed power available, it runs only 4 workers irrespective of the workload. In contrast, the dynamic policy can scale up to a higher number of workers to process a higher request rate. Figure~\ref{fig:leveraging-batteries}(c) shows the 95$^{th}$\%ile latency for the web application. The static (system-level) policy safeguards against the server going down, resulting in a much higher latency under high workload, while the dynamic policy is always able to meet the target latency SLO. 

\noindent{\bf Key Takeaway.} Our ecovisor enables applications to control virtual batteries to satisfy their application-specific requirements, e.g., low runtime versus low latency, compared to an application-agnostic system-level policy.

\subsubsection{Multi-tenancy.} Figure~\ref{fig:batteries_multi}(a) and \ref{fig:batteries_multi}(b) show the state of charge and actual charging/discharging patterns, respectively, for the virtual batteries allocated to each application. Both applications concurrently ran on a shared multi-tenant platform, but their battery usage patterns differ significantly depending on their requirements.

\begin{figure}[t]
    \centering
    \begin{tabular}{c}
    \includegraphics[width=0.97\columnwidth]{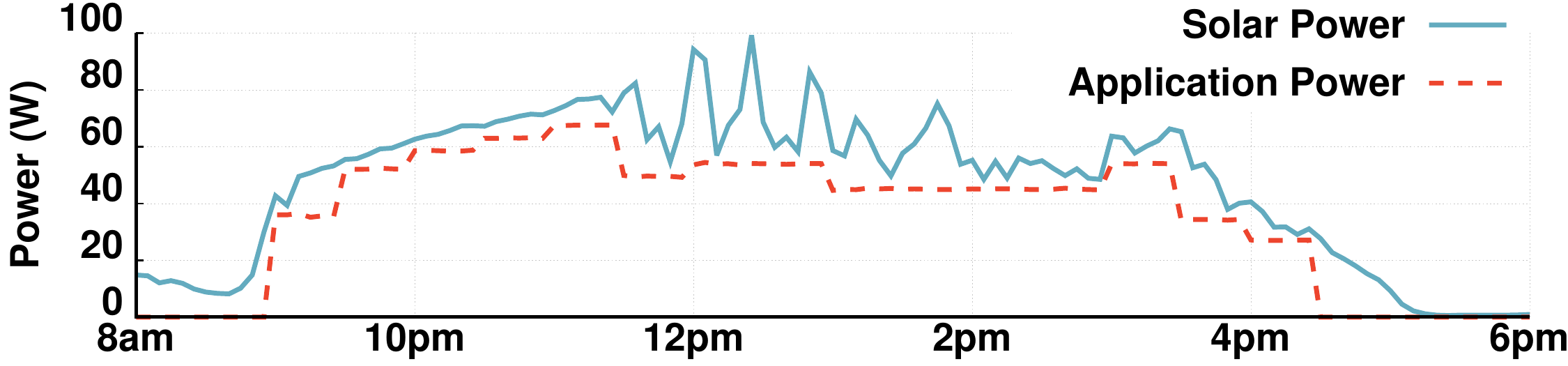}\\
    (a) Solar power\\
    \includegraphics[width=\columnwidth]{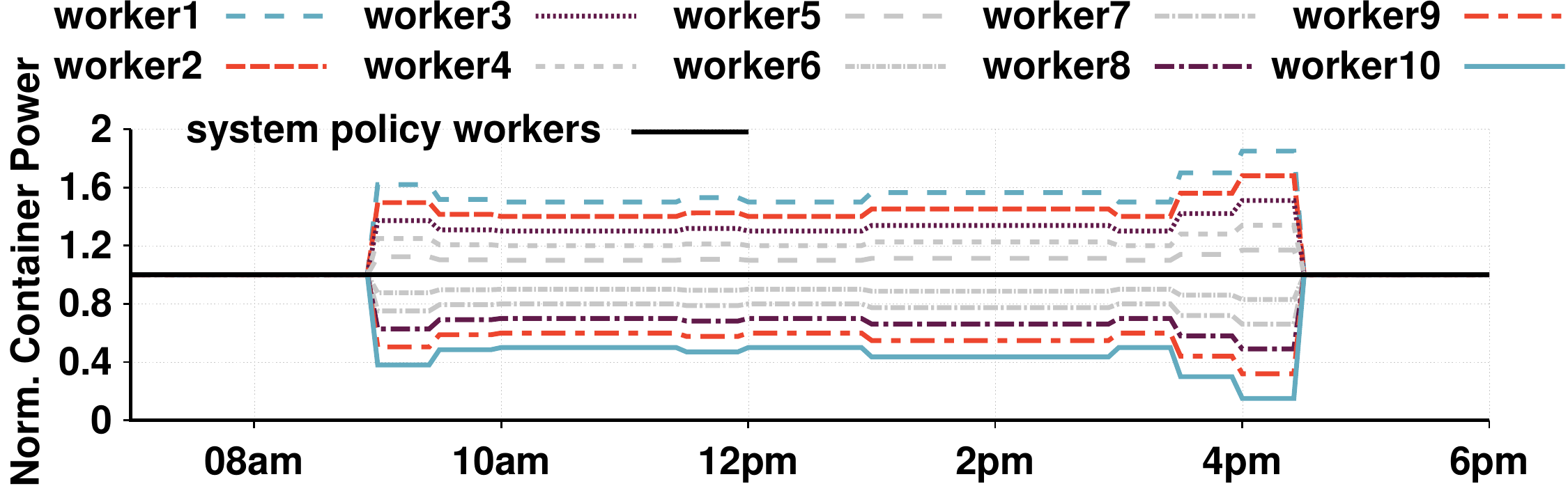}\\\vspace{0.15cm}
    (b) Per-container power caps\\
    \includegraphics[width=\columnwidth]{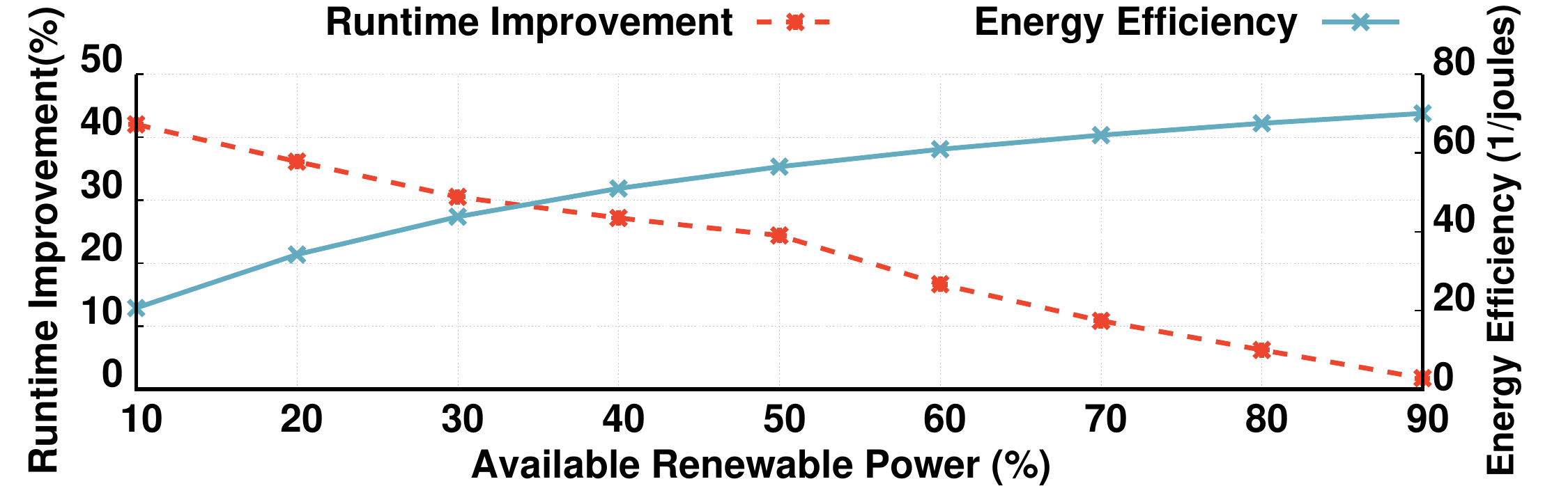}\\
    (c) Runtime and Energy-efficiency\\
    \end{tabular}
    \caption{\emph{Comparison of a static system-level and dynamic application-specific policy for setting per-container power caps (b) for a parallel job running across 10 nodes using the solar energy in (a). The dynamic policy has a better runtime and energy-efficiency (c).}}
    \label{fig:balancing-app}
\end{figure}

\subsection{Directly Exploiting Solar Power Efficiency}

Some parallel applications may more directly exploit solar power without any battery capacity, despite its volatility, using vertical scaling.   Since our ecovisor enables applications to balance power's supply and demand, these applications can explicitly allocate their limited solar power across a set of containers, e.g., such that the sum of containers' power caps does not exceed the supply of solar power. In this case, applications should allocate their limited solar power to where it can be used most productively.  
Since servers are not energy-proportional, they consume some power even when idle.  Thus, servers' most energy-efficient operating point is at 100\% of their allocated energy, and any idleness due to operating below this point wastes energy.
However, parallel applications often have tasks that are idle due to performing I/O, such as due to periodic task synchronization in the PyTorch training above. Such parallel applications frequently exhibit straggler tasks that increase running time by forcing other tasks to wait~\cite{straggler2,straggler1,straggler3}. 
Importantly, executing parallel applications on a limited amount of solar power can exacerbate the performance issues above.

\subsubsection{Applications.} To illustrate our policies for exploiting solar, we deploy two configurations of a synthetic parallel job.  In the first configuration, the job periodically synchronizes across tasks and performs I/O, using vertical scaling on all containers to match the available solar power.
In the second, we configure the parallel job to perform straggler mitigation by tracking the progress of each task, issuing a new replica for any slow task. For this configuration, we randomly inject straggler tasks in the workload.
We implement two power capping policies for the first configuration: (i) a system policy that sets static caps across 10 nodes, and (ii) an application policy that dynamically varies caps to ensure each node uses nearly all of their allocated energy, i.e., 100\% resource utilization.  Finally, the third policy handles stragglers by allocating extra resources when excess energy is available.

\subsubsection{Comparing Solar Policies.}
Figure~\ref{fig:balancing-app}(a) shows solar power availability for a single day.  In Figure~\ref{fig:balancing-app}(b), the dynamic power caps differ across the 10 nodes relative to the static cap (center line) over the trace.  Figure~\ref{fig:balancing-app}(c) scales the solar output from (a) by the percentage on the x-axis and plots the runtime improvement from using the dynamic policy (left y-axis) and its energy-efficiency (right y-axis).  The graph shows that as solar energy decreases, the importance of dynamically balancing power to reduce runtime increases.  Energy-efficiency increases as available solar power increases, since each node's base power is amortized over more productive work, which again illustrates the inefficiency of solar power. 

\begin{figure}[t]
    \centering
    \includegraphics[width=\columnwidth]{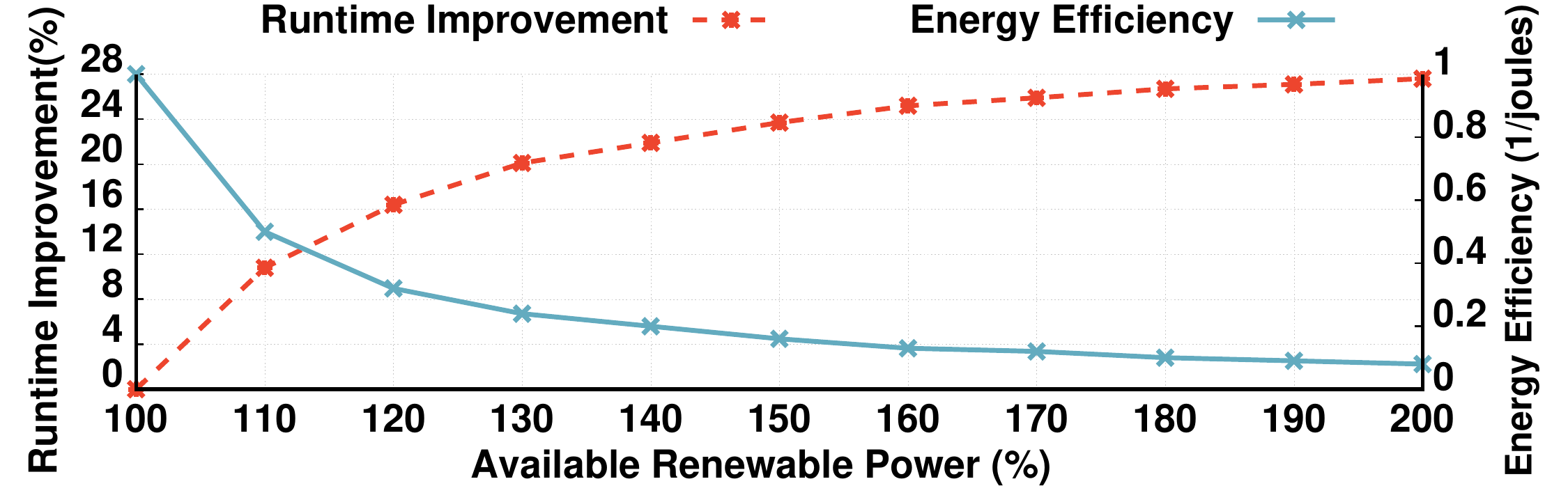}
    \caption{\emph{Mitigating stragglers using replica tasks is one way to productively consume excess renewable energy.}}
    \label{fig:straggler-app}
\end{figure} 

Finally, Figure~\ref{fig:straggler-app} shows the the third policy as we scale the solar output from Figure~\ref{fig:balancing-app}(a) up, which results in an excess of solar energy.   If applications cannot store the excess energy, they are incentivized to use it immediately, even if that usage is not entirely efficient. 
Figure~\ref{fig:straggler-app} shows that as solar energy increases, our application's overall energy-efficiency decreases, since we consume that energy by spawning more task replicas.   However, in this case, the absolute decrease in energy-efficiency is not important, since the excess solar energy would have been otherwise wasted. In this case, the application decreases its runtime by using the excess energy for straggler mitigation, although it sees diminishing returns as it submits more replicas (since at most one replica task will finish).  

\noindent{\bf \textit{Key Takeaway}.} Our ecovisor enables dynamic application-specific policies that improve solar-efficiency and performance using vertical scaling and straggler mitigation techniques compared to a system-level policy.

%% file: related.tex
Our work builds on prior work in managing energy, including integrating renewables and batteries into data centers, and recent work on managing carbon emissions.

\noindent {\bf Energy Management}.   There has been significant work on improving energy-efficiency and managing energy in computer systems over the past three decades.  This work has been highly successful in improving computing's energy-efficiency and reducing energy costs. Our work differs from this large class of work in its focus on carbon-efficiency, which is different from energy-efficiency.  Designing energy-efficient systems requires looking ``inward'' at various components to optimize their energy use, while designing carbon-efficient systems instead requires looking ``outward'' to the local energy system and grid to understand energy's source and characteristics.  

Our work is related to prior work that virtualizes power~\cite{deng2012,virtualpower,power-containers,virtual-utility} and exposes power management to applications~\cite{deng2012,chameleon,virtual-utility} across a variety of platforms.  For example, Nathuji and Schwan integrate hardware power management mechanisms with hypervisors to enable VM-level power management for servers~\cite{virtualpower}.  Similarly, Shen et al. propose power containers to enable fine-grained power management on a per-container basis. Our work leverages similar techniques for attributing and capping power for specific VMs, containers, and processes based on their resource usage~\cite{google-osdi20,capping-asplos}.  However, our work differs in its focus on using these mechanisms to expose visibility and control of the energy system itself, including power's carbon and availability characteristics, as well as control of a virtual battery.  This visibility and control enables applications to adapt their behavior to optimize carbon-efficiency in addition to energy-efficiency. 

Prior work has also proposed exposing power management to applications across a variety of platforms, including cloud platforms~\cite{deng2012, virtual-utility}, individual servers, and mobile devices~\cite{chameleon}.
Our work differs in that reducing power usage is not the same as reducing carbon emissions. 

\noindent {\bf Renewables and Storage Integration}.  There has also been significant prior work on integrating renewables and energy storage into data centers and optimizing applications for them.   Researchers have long recognized the potential to adapt cloud applications to variable renewable energy. Thus, prior work has optimized numerous applications, including Hadoop~\cite{greenhadoop}, job schedulers~\cite{greenslot}, key-value stores~\cite{greencassandra}, distributed storage~\cite{sharma:asplos11}, load balancers~\cite{gupta-solar}, and job migration~\cite{hotnets21} to run on variable renewable energy. While this prior work must implicitly embed ecovisor-like APIs within their systems and applications, they do not define and expose these APIs externally, and thus cannot support different application-level carbon and energy management policies.  GreenSwitch is perhaps most related to our ecovisor, as it defines a model-based policy for dynamically scheduling workload and selecting energy sources, e.g., grid, battery, solar, to optimize various objectives, e.g., cost, peak power, carbon footprint~\cite{parasol}. However, as above, GreenSwitch implements its policy at the system-level and does not expose visibility or control of the energy system to applications.

There has also been significant prior work on leveraging batteries in cloud data centers to reduce energy costs and provide power during outages~\cite{bhuvan1,dr2,bhuvan2}.  Our work provides applications the visibility and control necessary to actually implement these optimizations on a shared multi-tenant infrastructure, which prior work has not previously addressed, and also provides a platform for developing new optimizations, as we show in \Section\ref{sec:evaluation}. 

\noindent {\bf Carbon Management}.  Recent work has recognized the importance of reducing carbon emissions, and has attempted to quantify the carbon emissions of running particular applications~\cite{stochastic-parrots,deng2012, dean-carbon,mccallum}. Recent work has also attempted to quantify the carbon footprint of cloud data centers, including the carbon emissions embedded in hardware, i.e., Scope 3 emissions~\cite{facebook-carbon-neutral}.   However, this work does not provide any solutions for enabling carbon-efficiency optimizations.  As we show, carbon optimizations are now possible with the emergence of carbon information services, such as electricityMap~\cite{electricity-map}, which provide real-time estimates of grid power's carbon-intensity.  Recent work has proposed nascent carbon-efficiency optimizations, such as WaitAWhile~\cite{wait-awhile}.   Our ecovisor can enable these and other application-specific carbon-efficiency optimizations concurrently on a shared platform.

%% file: conclusion.tex
Enabling the design of carbon-efficient applications is an important research area that is necessary to halt climate change.  To address the problem, we propose an ecovisor that virtualizes the energy system and exposes software-defined visibility into, and control of, it to applications. Our approach pushes visibility and control of the energy system from hardware into software, enabling applications to optimize carbon-efficiency based on their own requirements. We build a small-scale ecovisor prototype, and demonstrate its ability to support a variety of carbon-efficiency optimizations for different applications.  
In the future, we plan to enable coordination between distributed ecovisor clusters to enable geo-distributed applications.

%% file: paper.bbl
\begin{thebibliography}{10}

\bibitem{amazon-carbon-neutral}
Reuters, {A}mazon {V}ows to be {C}arbon {N}eutral by 2040, buying 100,000
  {E}lectric {V}ans.
\newblock
  \url{https://www.reuters.com/article/us-amazon-environment/amazon-vows-to-be-carbon-neutral-by-2040-buying-100000-electric-vans-idUSKBN1W41ZV},
  September 19th 2019.

\bibitem{aws-spot}
Amazon {E}{C}2 {S}pot {I}nstances.
\newblock \url{https://aws.amazon.com/ec2/spot/}, Accessed March 2022.

\bibitem{aws-autoscale}
Aws autoscaling.
\newblock \url{https://aws.amazon.com/autoscaling/}, Accessed February 2022.

\bibitem{azure-monitor}
Azure {M}onitor.
\newblock \url{https://azure.microsoft.com/en-us/services/monitor/}, Accessed
  July 2022.

\bibitem{azure-spot}
Azure {S}pot {V}irtual {M}achines.
\newblock \url{https://azure.microsoft.com/en-us/pricing/spot/}, Accessed March
  2022.

\bibitem{blast}
{B}asic {L}ocal {A}lignment {S}earch {T}ool.
\newblock \url{https://blast.ncbi.nlm.nih.gov/Blast.cgi}, Accessed March 2022.

\bibitem{caiso}
California {I}{S}{O}.
\newblock \url{https://www.caiso.com/Pages/default.aspx}, Accessed March 2022.

\bibitem{google-carbon}
Carbon {F}ree {E}nergy for {G}oogle {C}loud {R}egions.
\newblock \url{https://cloud.google.com/sustainability/region-carbon}, Accessed
  March 2022.

\bibitem{electricity-map}
Electricity {M}ap.
\newblock \url{https://www.electricitymap.org/map}, Accessed March 2022.

\bibitem{google-carbon3}
Google cloud carbon footprint: Measure, report and reduce your cloud carbon
  emisisons.
\newblock \url{https://cloud.google.com/carbon-footprint}, Accessed July 2022.

\bibitem{google-pue}
Google {D}ata {C}enters {E}fficiency.
\newblock \url{google.com/about/datacenters/efficiency/}, Accessed March 2022.

\bibitem{ghg}
Greenhouse {G}as {P}rotocol.
\newblock \url{https://ghgprotocol.org/}, Accessed March 2022.

\bibitem{ipmi}
I{P}{M}{I} {O}verview.
\newblock
  \url{https://www.ibm.com/docs/en/power9/0009-ESS?topic=ipmi-overview},
  Accessed March 2022.

\bibitem{azure-autoscale}
Overview of autoscale in microsoft azure.
\newblock
  \url{https://docs.microsoft.com/en-us/azure/azure-monitor/autoscale/autoscale-overview},
  Accessed February 2022.

\bibitem{tesla-api}
Python {T}esla {P}owerwall {A}{P}{I}.
\newblock \url{https://pypi.org/project/tesla-powerwall/}, Accessed March 2022.

\bibitem{rock64-board}
R{O}{C}{K}64.
\newblock \url{https://www.pine64.org/devices/single-board-computers/rock64/},
  Accessed March 2022.

\bibitem{tesla-powerwall}
Tesla {P}owerwall {M}odes of {O}peration with {S}olar.
\newblock
  \url{https://www.tesla.com/en\_au/support/energy/powerwall/mobile-app/modes-of-operationwithsolar},
  Accessed March 2022.

\bibitem{uptime}
Uptime {I}nstitute {G}lobal {D}ata {C}enter {S}urvey 2021: Growth {S}tretches
  an {E}volving {S}ector.
\newblock
  \url{https://uptimeinstitute.com/resources/asset/2021-data-center-industry-survey},
  Accessed May 2022.

\bibitem{watttime}
Watt{T}ime.
\newblock \url{https://www.watttime.org/}, Accessed March 2022.

\bibitem{vmware-carbon}
Nicola Acutt.
\newblock {R}adius: Stories at the {E}dge, {A}chieving {C}arbon {N}eutrality.
\newblock \url{https://www.vmware.com/radius/achieving-carbon-neutrality/},
  November 1st 2018.

\bibitem{hotnets21}
Anup Agarwal, Jinghan Sun, Shadi Noghabi, Srinivasan Iyengar, Anirudh Badam,
  Ranveer Chandra, Srinivasan Seshan, and Shivkumar Kalyanaraman.
\newblock {R}edesigning {C}loud {C}omputing for {R}enewable {E}nergy.
\newblock In {\em Proceedings of the Twentieth ACM Workshop on Hot Topics in
  Networks (HotNets)}, November 2021.

\bibitem{ml-compute-demand}
Dario Amodei, Danny Hernandez, Girish Sastry, Jack Clark, Greg Brockman, and
  Ilya Sutskever.
\newblock A{I} and {C}ompute.
\newblock \url{https://openai.com/blog/ai-and-compute/}, May 16th 2018.

\bibitem{straggler2}
Ganesh Ananthanarayanan, Ali Ghodsi, Scott Shenker, and Ion Stoica.
\newblock Effective {S}traggler {M}itigation: Attack of the {C}lones.
\newblock In {\em USENIX Symposium on Networked System Design and
  Implementation (NSDI)}, April 2013.

\bibitem{straggler1}
Ganesh Ananthanarayanan, Michael Chien-Chun Hung, Xiaoqi Ren, Ion Stoica, Adam
  Wierman, and Minlan Yu.
\newblock G{R}{A}{S}{S}: Trimming {S}tragglers in {A}pproximation {A}nalytics.
\newblock In {\em USENIX Symposium on Networked System Design and
  Implementation (NSDI)}, April 2014.

\bibitem{stochastic-parrots}
Emily~M Bender, Timnit Gebru, Angelina McMillan-Major, and Shmargaret
  Shmitchell.
\newblock On the {D}angers of {S}tochastic {P}arrots: Can {L}anguage {M}odels
  {B}e {T}oo {B}ig?
\newblock In {\em Proceedings of the 2021 ACM Conference on Fairness,
  Accountability, and Transparency (FAccT)}, March 2021.

\bibitem{solarprotocol}
T.~Brain, A.~Nathanson, and B.~Piantella.
\newblock {S}olar {P}rotocol.
\newblock \url{http://solarprotocol.net/}, July 26th 2022.

\bibitem{power-api}
Maxime Colmant, Pascal Felber, Romain Rouvoy, and Lionel Seinturier.
\newblock Watts{K}it: Software-{D}efined {P}ower {M}onitoring of {D}istributed
  {S}ystems.
\newblock In {\em 17th IEEE/ACM International Symposium on Cluster, Cloud and
  Grid Computing (CCGRID)}, April 2017.

\bibitem{lowtechmagazine}
K.~De Decker.
\newblock This {W}ebsite {R}uns on a {S}olar {P}owered {S}erver {L}ocated in
  {B}arcelona.
\newblock \url{https://solar.lowtechmagazine.com/power.html}, July 26th 2022.

\bibitem{deng2012}
Nan Deng, Christopher Stewart, Daniel Gmach, and Martin Arlitt.
\newblock Policy and {M}echanism for {C}arbon-{A}ware {C}loud {A}pplications.
\newblock In {\em 2012 IEEE Network Operations and Management Symposium}, 2012.

\bibitem{exokernel}
Dawson~R. Engler, M.~Frans Kaashoek, and James O'Toole.
\newblock Exokernel: An {O}perating {S}ystem {A}rchitecture for
  {A}pplication-{L}evel {R}esource {M}anagement.
\newblock In {\em ACM Symposium on Operating System Principles (SOSP)}, 1995.

\bibitem{google-carbon-free}
Darrell Etherington.
\newblock Tech{C}runch, {G}oogle {C}laims {N}et {Z}ero {C}arbon {F}ootprint
  over its {E}ntire {L}ifetime, {A}ims to only use {C}arbon-{F}ree {E}nergy by
  2030.
\newblock
  \url{https://techcrunch.com/2020/09/14/google-claims-net-zero-carbon-footprint-over-its-entire-lifetime-aims-to-only-use-carbon-free-energy-by-2030/},
  September 14th 2020.

\bibitem{drf}
Ali Ghodsi, Matei Zaharia, Benjamin Hindman, Andy Konwinski, Scott Shenker, and
  Ion Stoica.
\newblock Dominant {R}esource {F}airness: Fair {A}llocation of {M}ultiple
  {R}esource {T}ypes.
\newblock In {\em USENIX Symposium on Networked System Design and
  Implementation (NSDI)}, April 2011.

\bibitem{greenslot}
Inigo Goiri, Ryan Beauchea, Kien Le, Thu~D. Nguyen, Md.~E. Haque, Jordi
  Guitart, Jordi Torres, and Ricardo Bianchini.
\newblock Green{S}lot: Scheduling {E}nergy {C}onsumption in {G}reen
  {D}atacenters.
\newblock In {\em ACM/IEEE International Conference for High Performance
  Computing, Networking, Storage and Analysis (SC)}, Seattle, Washington,
  November 2011.

\bibitem{parasol}
Inigo Goiri, William Katsak, Kien Le, Thu~D. Nguyen, and Ricardo Bianchini.
\newblock Parasol and {G}reen{S}witch: Managing {D}atacenters {P}owered by
  {R}enewable {E}nergy.
\newblock In {\em ACM Conference on Architectural Support for Programming
  Languages and Operating Systems (ASPLOS)}, March 2013.

\bibitem{greenhadoop}
Inigo Goiri, Kien Le, Thu~D. Nguyen, Jordi Guitart, Jordi Torres, and Ricardo
  Bianchini.
\newblock Green{H}adoop: Leveraging {G}reen {E}nergy in {D}ata-processing
  {F}rameworks.
\newblock In {\em ACM European Conference on Computer Systems (EuroSys)}, April
  2012.

\bibitem{bhuvan1}
Sriram Govindan, Anand Sivasubramaniam, and Bhuvan Urgaonkar.
\newblock Benefits and {L}imitations of {T}apping into {S}tored {E}nergy for
  {D}atacenters.
\newblock In {\em Proceedings of the 38th Annual International Symposium on
  Computer Architecture (ISCA)}, June 2011.

\bibitem{cloud-watch}
Developer Guide.
\newblock Amazon {C}loudwatch.
\newblock 2009.

\bibitem{gupta-solar}
Vani Gupta, Prashant Shenoy, and Ramesh~K Sitaraman.
\newblock Combining {R}enewable {S}olar and {O}pen {A}ir {C}ooling for
  {G}reening {I}nternet-{S}cale {D}istributed {N}etworks.
\newblock In {\em Proceedings of the Tenth ACM International Conference on
  Future Energy Systems (e-Energy)}, June 2019.

\bibitem{straggler3}
Aaron Harlap, Henggang Cui, Wei Dai, Jinliang Wei, Gregory Ganger, Phillip
  Gibbons, Garth Gibson, and Eric Xing.
\newblock Addressing the {S}traggler {P}roblem for {I}terative {C}onvergent
  {P}arallel {M}{L}.
\newblock In {\em Symposium on Cloud Computing (SoCC)}, September 2016.

\bibitem{guardian}
Fiona Harvey.
\newblock The {G}uardian, {M}ajor {C}limate {C}hanges {I}nevitable and
  {I}rreversible – {I}{P}{C}{C}'s {S}tarkest {W}arning {Y}et.
\newblock
  \url{https://www.theguardian.com/science/2021/aug/09/humans-have-caused-unprecedented-and-irreversible-change-to-climate-scientists-warn},
  August 9th 2021.

\bibitem{tigres}
Valerie Hendrix, James Fox, Devarshi Ghoshal, and Lavanya Ramakrishnan.
\newblock Tigres {W}orkflow {L}ibrary: Supporting {S}cientific {P}ipelines on
  {H}{P}{C} {S}ystems.
\newblock In {\em 16th IEEE/ACM International Symposium on Cluster, Cloud and
  Grid Computing (CCGRID)}, May 2016.

\bibitem{mesos}
Benjamin Hindman, Andy Konwinski, Matei Zaharia, Ali Ghodsi, Anthony Joseph,
  Randy Katz, Scott Shenker, and Ion Stoica.
\newblock Mesos: A {P}latform for {F}ine-grained {R}esource {S}haring in the
  {D}ata {C}enter.
\newblock In {\em USENIX Symposium on Networked Systems Design and
  Implementation (NSDI)}, March 2011.

\bibitem{carbon-offsets}
Umair Irfan.
\newblock Vox, {C}an {Y}ou {R}eally {N}egate {Y}our {C}arbon {E}missions?
  {C}arbon {O}ffsets, {E}xplained.
\newblock
  \url{https://www.vox.com/2020/2/27/20994118/carbon-offset/-climate-change-net-zero-neutral-emissions},
  February 27th 2020.

\bibitem{maiden}
Penny Jones.
\newblock Data{C}enter{D}ynamics, {A}pple {C}onfirms {S}olar {F}arm at {M}aiden
  {D}ata {C}enter.
\newblock
  \url{https://www.datacenterdynamics.com/en/news/apple-confirms-solar-farm-at-maiden-data-center/},
  February 21st 2012.

\bibitem{greencassandra}
William Katsak, Inigo Goiri, Ricardo Bianchini, and Thu Nguyen.
\newblock Green{C}assandra: Using {R}enewable {E}nergy in {D}istributed
  {S}tructured {S}torage {S}ystems.
\newblock In {\em International Conference on Green and Sustainable Computing
  (IGSC)}, June 2015.

\bibitem{resnet34}
Brett Koonce.
\newblock Resnet 34.
\newblock In {\em Convolutional {N}eural {N}etworks with {S}wift for
  {T}ensorflow}. Springer, 2021.

\bibitem{cifar100}
Alex Krizhevsky, Geoffrey Hinton, et~al.
\newblock Learning {M}ultiple {L}ayers of {F}eatures from {T}iny {I}mages.
\newblock 2009.

\bibitem{google-osdi20}
Shaohong Li, Xi~Wang, Faria Kalim, Xiao Zhang, Sangeetha~Abdu Jyothi, Karan
  Grover, Vasileios Kontorinis, Nina Narodytska, Owolabi Legunsen, Sreekumar
  Kodakara, et~al.
\newblock Thunderbolt: Throughput-{O}ptimized, {Q}uality-of-{S}ervice-{A}ware
  {P}ower {C}apping at {S}cale.
\newblock In {\em USENIX Symposium on Operating System Design and
  Implementation (OSDI)}, November 2020.

\bibitem{chameleon}
Xiaotao Liu, Prashant Shenoy, and Mark~D Corner.
\newblock Chameleon: {A}pplication-level {P}ower {M}anagement.
\newblock {\em IEEE Transactions on Mobile Computing}, 2008.

\bibitem{lxd}
Canonical Ltd.
\newblock L{X}{D}.
\newblock \url{https://linuxcontainers.org/lxd/introduction/}.

\bibitem{masanet}
Eric Masanet, Arman Shehabi, Nuoa Lei, Sarah Smith, and Jonathan Koomey.
\newblock Recalibrating {G}lobal {D}ata {C}enter {E}nergy-use {E}stimates.
\newblock {\em Science}, 367(6481):984--986, February 2020.

\bibitem{ipcc-report}
Val{\'e}rie Masson-Delmotte, Panmao Zhai, Anna Pirani, Sarah~L Connors,
  Clotilde P{\'e}an, Sophie Berger, Nada Caud, Yang Chen, Leah Goldfarb,
  Melissa~I Gomis, et~al.
\newblock Summary for {P}olicymakers. in: {C}limate {C}hange 2021: The
  {P}hysical {S}cience {B}asis. {C}ontribution of {W}orking {G}roup {I} to the
  {S}ixth {A}ssessment {R}eport of the {I}ntergovernmental {P}anel on {C}limate
  {C}hange.
\newblock Technical report, United Nation Intergovernmental Panel on Climate
  Change (IPCC), 2021.

\bibitem{virtualpower}
Ripal Nathuji and Karsten Schwan.
\newblock Virtual{P}ower: Coordinated {P}ower {M}anagement in {V}irtualized
  {E}nterprise {S}ystems.
\newblock In {\em ACM Symposium on Operating System Principles (SOSP)}, October
  2007.

\bibitem{facebook-carbon-neutral}
Kevin O'Sullivan.
\newblock The {I}rish {T}imes, {F}acebook {C}ommits to {N}et-{Z}ero {C}arbon
  {E}missions by 2030.
\newblock
  \url{https://www.irishtimes.com/news/environment/facebook-commits-to-net-zero-carbon-emissions-by-2030-1.4354701},
  September 15th 2020.

\bibitem{dr2}
Darshan~S. Palasamudram, Ramesh~K. Sitaraman, Bhuvan Urgaonkar, and Rahul
  Urgaonkar.
\newblock Using {B}atteries to {R}educe {P}ower {C}osts of {I}nternet-{S}cale
  {D}istributed {N}etworks.
\newblock In {\em ACM Symposium on Cloud Computing (SoCC)}, October 2012.

\bibitem{dcd-article}
Andy Patrizio.
\newblock Data {C}enter {D}ynamics, {T}he {Q}uest to {R}un {D}ata {C}enters on
  {B}attery {P}ower.
\newblock
  \url{https://www.datacenterdynamics.com/en/analysis/the-quest-to-run-data-centers-on-battery-power/},
  April 5th 2021.

\bibitem{dean-carbon}
David Patterson, Joseph Gonzalez, Quoc Le, Chen Liang, Lluis-Miquel Munguia,
  Daniel Rothchild, David So, Maud Texier, and Jeff Dean.
\newblock Carbon {E}missions and {L}arge {N}eural {N}etwork {T}raining.
\newblock Technical report, arXiv, April 2021.

\bibitem{google-blog2}
Sundar Pichai.
\newblock Google {B}log, {N}ew progress toward our 24/7 carbon-free energy
  goal.
\newblock
  \url{https://blog.google/outreach-initiatives/sustainability/new-progress-toward-our-247-carbon-free-energy-goal},
  April 20th 2021.

\bibitem{microsoft}
John Roach.
\newblock Microsoft datacenter batteries to support growth of renewables on the
  power grid.
\newblock
  \url{https://news.microsoft.com/innovation-stories/ireland-wind-farm-datacenter-ups/},
  July 7th 2022.

\bibitem{capping-asplos}
Varun Sakalkar, Vasileios Kontorinis, David Landhuis, Shaohong Li, Darren
  De~Ronde, Thomas Blooming, Anand Ramesh, James Kennedy, Christopher Malone,
  Jimmy Clidaras, et~al.
\newblock Data {C}enter {P}ower {O}versubscription with a {M}edium {V}oltage
  {P}ower {P}lane and {P}riority-{A}ware {C}apping.
\newblock In {\em ACM Symposium on Architectural Support for Programming
  Languages and Operating Systems (ASPLOS)}, March 2020.

\bibitem{endtoend}
Jerome~H. Saltzer, David~P. Reed, and David~D. Clark.
\newblock End-{T}o-{E}nd {A}rguments in {S}ystem {D}esign.
\newblock {\em ACM Transactions on Computer Systems}, November 1984.

\bibitem{sharma:asplos11}
Navin Sharma, Sean Barker, David Irwin, and Prashant Shenoy.
\newblock {B}link: {M}anaging {S}erver {C}lusters on {I}ntermittent {P}ower.
\newblock In {\em ACM Conference on Architectural Support for Programming
  Languages and Operating Systems (ASPLOS)}, March 2011.

\bibitem{power-containers}
Kai Shen, Arrvindh Shriraman, Sandhya Dwarkadas, Xiao Zhang, and Zhuan Chen.
\newblock Power {C}ontainers: An {O}{S} {F}acility for {F}ine-grained {P}ower
  and {E}nergy {M}anagement on {M}ulticore {S}ervers.
\newblock In {\em ACM Conference on Architectural Support for Programming
  Languages and Operating Systems (ASPLOS)}, March 2013.

\bibitem{yank}
Rahul Singh, David Irwin, Prashant Shenoy, and K.K. Ramakrishnan.
\newblock Yank: Enabling {G}reen {D}ata {C}enters to {P}ull the {P}lug.
\newblock In {\em USENIX Symposium on Networked Systems Design and
  Implementation (NSDI)}, April 2013.

\bibitem{microsoft-carbon-negative}
Brad Smith.
\newblock Official {M}icrosoft {B}log, {M}icrosoft will be {C}arbon {N}egative
  by 2030.
\newblock
  \url{https://blogs.microsoft.com/blog/2020/01/16/microsoft-will-be-carbon-negative-by-2030/},
  January 16th 2020.

\bibitem{mccallum}
Emma Strubell, Ananya Ganesh, and Andrew Mc{C}allum.
\newblock Energy and {P}olicy {C}onsiderations for {M}odern {D}eep {L}earning
  {R}esearch.
\newblock In {\em AAAI Conference on Artificial Intelligence (AAAI)}, February
  2020.

\bibitem{wikipedia-trace}
Guido Urdaneta, Guillaume Pierre, and Maarten van Steen.
\newblock Wikipedia workload analysis for decentralized hosting.
\newblock {\em Elsevier Computer Networks}, July 2009.
\newblock \url{http://www.globule.org/publi/WWADH_comnet2009.html}.

\bibitem{bhuvan2}
Rahul Urgaonkar, Bhuvan Urgaonkar, Michael~J Neely, and Anand Sivasubramaniam.
\newblock {O}ptimal {P}ower {C}ost {M}anagement {U}sing {S}tored {E}nergy in
  {D}ata {C}enters.
\newblock In {\em Proceedings of the ACM Conference on Measurement and Analysis
  of Computing Systems (SIGMETRICS)}, March 2011.

\bibitem{virtual-utility}
Cheng Wang, Bhuvan Urgaonkar, George Kesidis, Uday~V Shanbhag, and Qian Wang.
\newblock A {C}ase for {V}irtualizing the {E}lectric {U}tility in {C}loud
  {D}ata {C}enters.
\newblock In {\em USENIX Workshop on Hot Topics in Cloud Computing (HotCloud)},
  July 2014.

\bibitem{wait-awhile}
Philipp Wiesner, Ilja Behnke, Dominik Scheinert, Kordian Gontarska, and Lauritz
  Thamsen.
\newblock Let's {W}ait {A}while: How {T}emporal {W}orkload {S}hifting {C}an
  {R}educe {C}arbon {E}missions in the {C}loud.
\newblock In {\em Proceedings of the 22nd International Middleware Conference
  (Middleware)}, December 2021.

\end{thebibliography}
